\documentclass[letterpaper,twocolumn,showpacs,preprintnumbers,10pt]{revtex4-1}

\usepackage{graphicx}
\usepackage{amsmath,amssymb}
\usepackage{dcolumn}
\usepackage{bm}

\usepackage{amsfonts}

\begin{document}

\preprint{APS/123-QED}

\title{Elasticidad de las membranas biol\'ogicas\\}

\author{L. Mart\'inez-Balbuena}
\affiliation{%
Departamento de Investigaci\'on en F\'isica, Universidad de Sonora,  
Apdo. Postal 5-088, 83190 Hermosillo, Sonora, M\'exico.\\ 
}%

\author{A. Maldonado-Arce}
\affiliation{
Departamento de F\'isica, Universidad de Sonora, Apdo. Postal 1626, 83000 Hermosillo, Sonora, M\'exico.\\
}%
\author{E. Hern\'andez-Zapata}
\affiliation{
Departamento de F\'isica, Matem\'aticas e Ingenier\'ia, Universidad de Sonora, C.P. 83600, H. Caborca, Sonora, M\'exico.\\
}%
\date{\today}

\begin{abstract}
En este art\'iculo se presenta una revisi\'on del conocimiento actual sobre las propiedades el\'asticas 
de las membranas de origen biol\'ogico. Bajo la hip\'otesis del mosaico fluido se considera a una bicapa de fosfol\'ipido 
como la base estructural de una membrana biol\'ogica y se expone un modelo debido a Helfrich a partir del cu\'al se obtiene una 
expresi\'on para la energ\'ia el\'astica de una membrana. La energ\'ia libre queda completamente caracterizada por sus curvaturas 
locales principales y cuatro par\'ametros dependientes de la composici\'on qu\'imica de la membrana y de su entorno local. 
Asimismo, se expone la justificaci\'on te\'orica de un m\'etodo experimental (que hace uso de micromanipulaci\'on con 
micropipetas) que permite medir la constante el\'astica de curvatura media de una membrana. Finalmente se describen algunas 
consecuencias f\'isicas de estos conceptos, como la explicaci\'on de la transici\'on de una fase lamelar a una esponja, o la 
aparici\'on de la interacci\'on est\'erica entre membranas.\\ 
\textit{Descriptores:} Energ\'ia libre de Helfrich; membranas biol\'ogicas; bicapas lip\'idicas; elasticidad de membranas.
\vspace{0.3cm}

In this article we review the present knowledge on the elastic properties of membranes of biological origin. Assuming the 
fluid mosaic hypothesis, we consider the phospholipid bilayer as the structural base of a biological membrane. We expose a 
model due to Helfrich which can be used to obtain an expression for the membrane elastic energy. The free energy is completely 
characterized by two local principal curvatures and four parameters that depend on the chemical composition of the membrane 
and its local environment. In addition we present the theoretical justification of an experimental method (the Micropipette 
Manipulation technique) that may be used to measure the membrane rigidity (the elastic constant that corresponds to the 
mean curvature of the membrane). Finally, we describe some physical consequences of these concepts, such as the explanation 
of the lamellar-sponge phase transition and the emergence of the steric interaction between membranes.
\end{abstract}

\maketitle

\section{\label{sec:level1}Introducci\'on}

Las membranas biol\'ogicas son componentes primordiales de todos los organismos \cite{yeagle, lodish}. Se trata de estructuras 
esencialmente formadas por una bicapa de fosfol\'ipido con prote\'inas incorporadas, que est\'an presentes en las c\'elulas 
biol\'ogicas as\'i como en los diferentes organelos. 

Las membranas tienen diferentes funciones biol\'ogicas. Por una parte, a\'islan a la c\'elula misma y a ciertos organelos, 
como el n\'ucleo celular, de su exterior. Pero adem\'as de servir como fronteras entre los espacios intra y extracelular o 
inter-extra organelo, las membranas tambi\'en participan en otras funciones como el transporte activo, el encapsulamiento 
de sustancias, el soporte de canales i\'onicos y la estabilidad de la propia c\'elula. Las membranas tambi\'en juegan un 
papel esencial en procesos biol\'ogicos como la endocitosis, la exocitosis, la fusi\'on celular y la formaci\'on de 
protuberancias \cite{yeagle, lodish}.

An\'alisis qu\'imicos y estudios biol\'ogicos han permitido conocer la composici\'on qu\'imica, as\'i como ciertas funciones 
biol\'ogicas de las membranas. Sin embargo, existen a\'un preguntas sin resolver en este campo de la Ciencia. Por ejemplo: 
\textquestiondown Cu\'ales son las propiedades f\'isicas relevantes de las membranas biol\'ogicas? \textquestiondown C\'omo 
dichas propiedades determinan o influencian la estructura y la funci\'on biol\'ogica de estos agregados? Entre las propiedades 
f\'isicas que reciben especial atenci\'on en estos tiempos se encuentran las din\'amicas, estructurales y mec\'anicas. En 
particular, entre las propiedades mec\'anicas destaca la elasticidad de las membranas, fen\'omeno que es\-t\'a relacionado 
de una manera apenas parcialmente conocida por el momento con fen\'omenos como la endocitosis y exocitosis, y que juega un 
papel relevante en la forma que adquieren algunos organelos en la c\'elula (ret\'iculo endopl\'asmico, complejo de Golgi).

En este trabajo, nos proponemos comentar algunos conceptos importantes para entender ciertas propiedades el\'asticas de 
las membranas. Prestaremos especial aten\-ci\'on a la teor\'ia de Wolfgang Helfrich, cient\'ifico alem\'an que lleg\'o a la 
expresi\'on utilizada actualmente para la energ\'ia el\'astica de una membrana.
\begin{figure}
 \centering
 \includegraphics[width=0.40\textwidth]{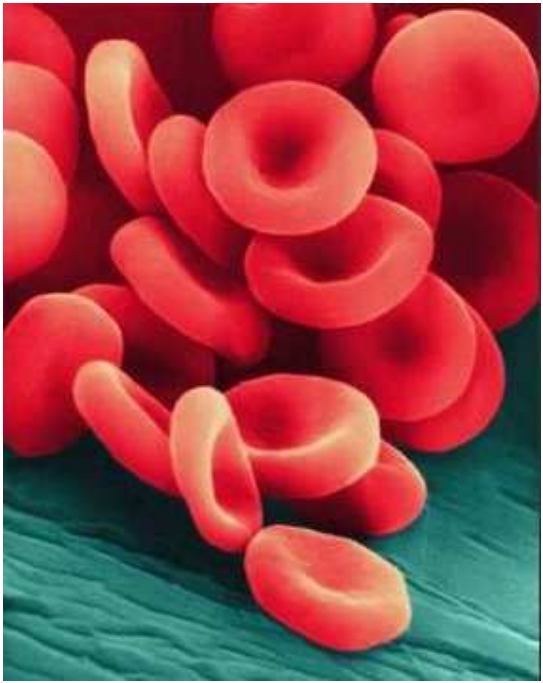}
  \caption{{\small Forma que los gl\'obulos rojos adoptan en su estado natural. Note como la membrana que rodea a 
cada c\'elula se tiene que curvar para adecuarse a la forma celular.}}
 \label{fig:rojos}
\end{figure}

Hist\'oricamente, el estudio de las propiedades el\'asticas de membranas est\'a relacionado con el problema de predecir 
te\'oricamente la forma de los gl\'obulos rojos (Figura \ref{fig:rojos}). Estas c\'elulas, y por tanto la membrana que las 
rodea, tienen forma discoidal, con una concavidad caracter\'istica en sus caras. En la b\'usqueda de entender el origen de 
esta estructura, en 1970 Canham \cite{canham} propuso un modelo en el que se considera constante tanto el \'area como el 
espesor de la membrana celular. En dicho modelo, los gl\'obulos rojos est\'an caracterizados \'unicamente por su elasticidad 
respecto a la curvatura, es decir, mediante una energ\'ia el\'astica de curvatura que, por unidad de \'area, adquiere la 
siguiente forma:
	\begin{equation}
	f=\frac{1}{2}\kappa H^{2}
	\end{equation}
donde $\kappa$ representaba el m\'odulo el\'astico de curvatura media de la membrana y $H$ la curvatura media de \'esta 
(ambas cantidades ser\'an definidas en la secci\'on \ref{sec:conceptos}). Veremos m\'as adelante que esta expresi\'on es parcialmente 
correcta. Sin embargo, este modelo conduce a formas no observadas en los gl\'obulos rojos \cite{deuling1}.

En la misma d\'ecada de los setentas, la comunidad de cient\'ificos de la materia condensada aprovech\'o las teor\'ias 
relativamente maduras de la f\'isica de cristales l\'iquidos termotr\'opicos para aplicarlas al estudio de membranas. Esto 
es posible debido a que localmente una membrana se comporta como una celda de cristal l\'iquido nem\'atico en alineamiento 
homeotr\'opico; esto es, mol\'eculas con forma de \textquotedblleft rodillos\textquotedblright alineadas perpendicularmente 
al sustrato. En este tipo de materiales, las mol\'eculas tienen un orden orientacional de largo alcance.
 
Aprovechando estos hechos, Helfrich utiliz\'o la expresi\'on para la densidad de ener\-g\'ia libre de Frank para cristales 
l\'iquidos uniaxiales \cite{frank}, considerando la normal a la membrana como el director del cristal l\'iquido. As\'i, 
dedujo la energ\'ia de curvatura  por unidad de \'area de una membrana \cite{helfrich}. La ecuaci\'on resultante, publicada 
en 1973, es llamada "energ\'ia libre de Helfrich" para membranas fluidas y es la base del an\'alisis de las propiedades 
el\'asticas de las membranas. En este trabajo presentaremos una deducci\'on alternativa de la energ\'ia de Helfrich y 
comentaremos algunas de sus consecuencias f\'isicas.

Nuestro art\'iculo est\'a dividido de la siguiente manera. En la secci\'on \ref{sec:bicapas}, se comentar\'a brevemente la 
ubicaci\'on biol\'ogica, as\'i como la estructura qu\'imica de las membranas celulares. En la secci\'on \ref{sec:helfrich} 
se expone el modelo de Helfrich para biomembranas. En las secciones siguientes se exponen algunas de las consecuencias f\'isicas 
de la elasticidad de curvatura: fluctuaciones t\'ermicas de membranas (secci\'on \ref{sec:forma}), m\'etodos de 
medici\'on de las constantes el\'asticas (\ref{sec:experimental}), transformaci\'on de la fase lamelar a la fase esponja 
(\ref{sec:transicion}) e interacciones est\'ericas entre membranas (\ref{sec:esterica}). Finalmente, en la secci\'on final damos 
algunas conclusiones del trabajo.

\subsection{\label{sec:bicapas}Agregados y bicapas lip\'idicas}

\begin{figure*}
	\centering
		\includegraphics[width=0.95\textwidth]{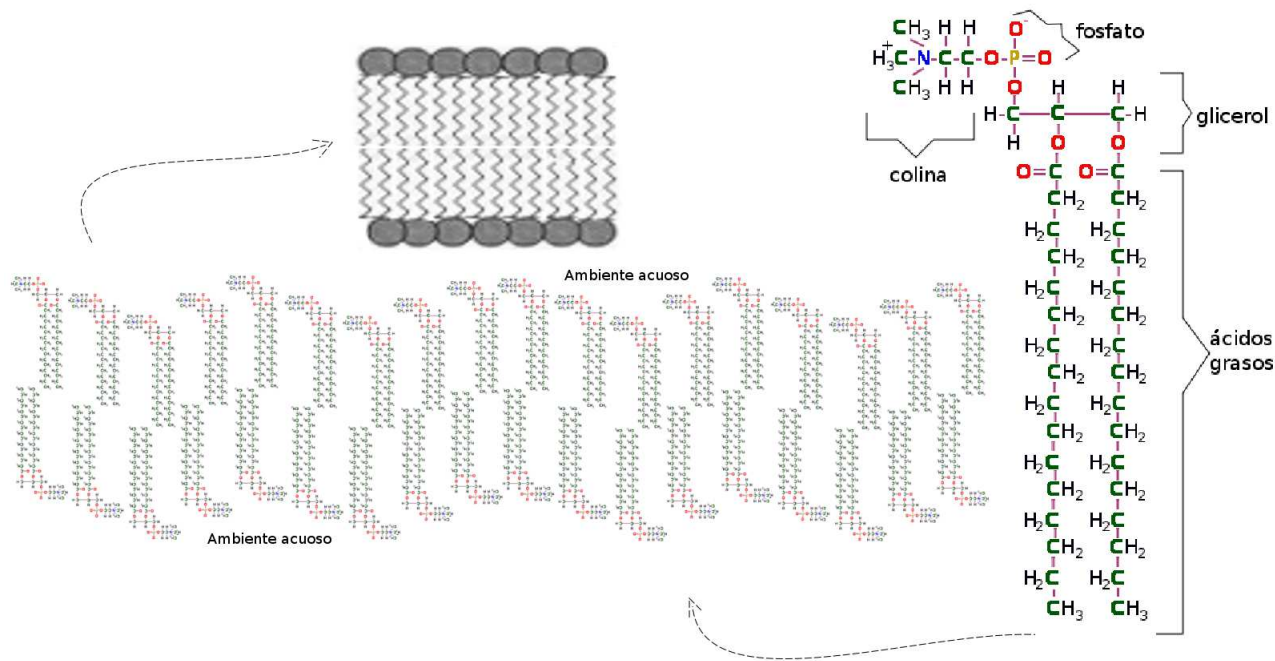}
	\caption{{\small Una de las componentes principales de una membrana biol\'ogica es una matriz l\'ipidica, en la cual, 
las colas hidrof\'obicas de las mol\'eculas de fosfol\'ipido evitan el contacto con el agua. En la figura mostramos tambi\'en 
la mol\'ecula de uno de los fosfol\'ipidos m\'as  comunes, de la familia de la fosfatidilcolina.}}
	\label{fig:acidosg}
\end{figure*}

Las membranas biol\'ogicas son estructuras muy complejas formadas por una gran variedad de mol\'eculas: fosfol\'ipidos, 
prote\'inas, etc. \cite{yeagle}. Estudios basados en micros\-cop\'ia electr\'onica y en el an\'alisis de la composici\'on 
qu\'imica de biomembranas, as\'i como estudios de propiedades f\'isicas tales como la permeabilidad y la difusi\'on de 
prote\'inas y de mol\'eculas lip\'idicas en membranas, condujeron al desarrollo del llamado modelo del mosaico fluido. 
Este modelo, propuesto por S. J. Singer y G. Nicolson en 1972 \cite{singer}, es el m\'as aceptado para describir a las 
membranas biol\'ogicas. La estructura base de la membrana es una bicapa lip\'idica donde las mol\'eculas individuales pueden 
moverse como en un l\'iquido bidimensional. La bicapa lip\'idica es una mezcla de varias clases de mol\'eculas, en particular 
fosfol\'ipidos y glicol\'ipidos, m\'as otras mol\'eculas peque\~nas como colesterol. Las prote\'inas de membrana se encuentran 
incrustadas en la bicapa lip\'idica. La estructura de las membranas biol\'ogicas es posible debido a que los fosfol\'ipidos 
pertenecen a un tipo especial de mol\'eculas, llamadas \textquotedblleft anfif\'ilicas\textquotedblright, a la cual tambi\'en 
pertenecen los surfactantes. Este nombre se debe a que una parte de la mol\'ecula (la cabeza polar) es soluble en agua, 
mientras que la otra parte (la cola hidrof\'obica) es insoluble en dicho solvente. As\'i, cuando estas mol\'eculas se 
disuelven en agua, las partes hidrof\'obicas tienden a asociarse para evitar el contacto con el solvente, donde son insolubles, 
lo cual da lugar a la estructura en forma de bicapas (recordemos que el medio celular es esencialmente acuoso). Las prote\'inas 
que se pueden insertar en dicha bicapa, tienen segmentos hidrof\'obicos que tampoco pueden solubilizarse en agua, los cuales 
se asocian con el centro hidrof\'obico de la membrana. En la Figura \ref{fig:acidosg} se muestra de forma esquem\'atica la 
estructura de una bicapa lip\'idica, as\'i como la f\'ormula molecular de uno de los fosfol\'ipidos m\'as conocidos, de la 
familia de las fosfatidilcolinas.

Mediante diferentes t\'ecnicas experimentales, como la dispersi\'on de radiaci\'on \cite{roux, caille, gennes} y la 
microscop\'ia electr\'onica \cite{yeagle}, se ha podido determinar el espesor de una membrana formada por mol\'eculas 
anfif\'ilicas (alrededor de 50\AA). Dicho espesor experimental coincide bastante bien con lo esperado a partir de la 
longitud de las mol\'eculas que forman las membranas.

Es importante subrayar que algunas propiedades f\'isicas (mec\'anicas y termodin\'amicas) de las membranas de fosfol\'ipido 
dependen crucialmente del tipo de l\'ipido. Esto se refleja principalmente en  la denominada temperatura de transici\'on, $T_{m}$, 
de los fosfol\'ipidos. Por debajo de esta temperatura, las mol\'eculas del fosfol\'ipido adquieren una estructura tipo gel 
(s\'olido amorfo bidimensional). Para temperaturas mayores que $T_{m}$, la membrana es fluida. Experimentos de calorimetr\'ia 
han mostrado c\'omo $T_{m}$ depende de las caracter\'isticas qu\'imicas de los l\'ipidos \cite{handbook}. Por ejemplo, para una cabeza 
polar dada (como fosfatidilcolina), $T_{m}$ se incrementa conforme se incrementa el n\'umero de \'atomos de carbono en la cadena 
hidrof\'obica. Esto se debe a que las cadenas de longitud grande magnifican el efecto hidrof\'obico, dando como resultado mayor 
cohesi\'on entre las mol\'eculas que forman la membrana. De manera similar, la presencia de insaturaciones (dobles enlaces qu\'imicos) 
en las colas hidrof\'obicas hace que $T_{m}$ disminuya, efecto que depende tambi\'en  de la posici\'on en que se presenta la 
insaturaci\'on \cite{handbook}. Se ha observado tambi\'en que la presencia de cierta asimetr\'ia entre las dos cadenas que forman la 
cola hidrof\'obica de la mol\'ecula, hacen que $T_{m}$ disminuya \cite{handbook}. Esto tiene como consecuencia que los fosfol\'ipidos m\'as 
utilizados para preparar modelos experimentales de membranas biol\'ogicas sean aquellos con una cadena saturada y otra insaturada, 
como es el caso del 1-estearoil-2-oleil-glicero 3-fosfocolina (SOPC), o de mezclas como la lecitina de soya o huevo. Esto asegura 
que la temperatura de transici\'on, $T_{m}$, est\'e siempre por debajo de la temperatura a la cual se llevan a cabo los experimentos.

\begin{figure}
 \centering
   \includegraphics[width=0.35\textwidth]{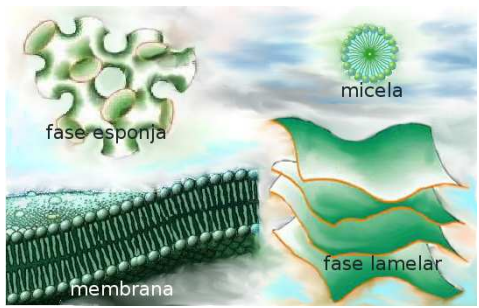}
   \caption{{\small Representaci\'on de la estructura de algunas de las fases t\'ipicas formadas por mol\'eculas anfif\'ilicas. 
La estructura m\'as sencilla es la micela. Por su parte, las membranas pueden ordenarse de maneras diferentes, tal como en la 
fase lamelar, en la esponja o en la vesicular.}}
   \label{fig:resumen}
\end{figure}

Cabe se\~nalar que las mol\'eculas anfif\'ilicas no solamente son capaces de formar las membranas biol\'ogicas, sino 
tambi\'en otro tipo de estructuras. Cuando estas mol\'eculas (ya sean fosfol\'ipidos o surfactantes) se disuelven en 
agua, su caracter\'istica dual en cuanto a  la solubilidad hace que se puedan formar diferentes estructuras, como las micelas 
(agregados individuales que pueden ser esf\'ericos o cil\'indricos) o las bicapas. A su vez, las bicapas, que son la estructura 
b\'asica de una membrana, pueden adoptar configuraciones en forma de ves\'iculas (agregado esf\'erico), fase lamelar (membranas 
paralelas), fase c\'ubica (membranas en arreglo peri\'odico) o fases esponja (bicapas desordenadas pero conectadas). Una 
representaci\'on esquem\'atica de algunas de estas fases se muestra en la Figura \ref{fig:resumen}.

Cada mol\'ecula particular presenta algunas de estas fases en diferentes condiciones fisicoqu\'imicas: concentraci\'on, 
temperatura, aditivos, etc. Por otra parte, se han desarrollado algunos modelos te\'oricos que predicen la topolog\'ia de 
los agregados. El m\'as sencillo de dichos modelos, propuesto por Israelachvili \cite{israelachvili}, estipula que la forma 
del agregado est\'a determinada por el llamado ``par\'ametro de empaquetamiento'', el cual se relaciona con la geometr\'ia 
de la mol\'ecula. Seg\'un este modelo simple, la asociaci\'on de mol\'eculas de forma c\'onica da lugar a un agregado con 
gran curvatura, como las micelas. En cambio, la asociaci\'on de mol\'eculas relativamente cil\'indricas da lugar a agregados 
con poca curvatura, como las membranas. Este modelo permite explicar algunas transiciones observadas en las fases de mol\'eculas 
anfif\'ilicas. 
Por ejemplo, si se tiene un surfactante cargado que forme micelas esf\'ericas, esto es debido a la forma efectivamente c\'onica 
de la mol\'ecula, magnificada por la repulsi\'on electrost\'atica entre cabezas polares. Al agregar sal al sistema, las 
interacciones electrost\'aticas se apantallan, lo cual tiene el efecto de disminuir el tama\~no efectivo lateral de las cabezas 
polares, por lo que la mol\'ecula se asemeja m\'as a un cilindro que a un cono. En estas condiciones, la teor\'ia del par\'ametro 
de empaquetamiento predice que la estructura que forman dichas mol\'eculas evoluciona de la forma micelar hacia la forma de membrana 
con el s\'olo hecho de agregar sal al sistema. Este efecto ha sido observado experimentalmente.

Sin embargo, hay otras transiciones que no han podido ser explicadas completamente mediante modelos geom\'etricos tan sencillos. 
Una de ellas es la transici\'on de la fase lamelar a la fase esponja (Figura \ref{fig:resumen}), donde membranas que tienen una 
configuraci\'on relativamente plana, adquieren una topolog\'ia local en forma de silla de montar a caballo. Se trata de una 
transformaci\'on topol\'ogica de las membranas. Lo mismo puede decirse en cambios donde se pasa de una fase lamelar a una fase 
vesicular, mediante la adici\'on de un pol\'imero por ejemplo, lo cual tambi\'en implica un cambio topol\'ogico. 
Veremos que este tipo de transformaciones se pueden entender en t\'erminos de la energ\'ia el\'astica de membranas, concepto que 
se discutir\'a en la siguiente secci\'on.

\section{\label{sec:helfrich}Modelo de Helfrich}

Algunas propiedades geom\'etricas y termodin\'amicas de las bicapas anfif\'ilicas pueden ser entendidas desde un punto de 
vista unificado considerando la energ\'ia libre asociada a sus deformaciones. El espesor tan peque\~no de una membrana (50\AA), 
al menos comparado con las dimensiones de una c\'elula (del orden de micras), permite aproximar a la bicapa como una superficie 
bidimensional inmersa en el solvente \cite{yang}. 

La membrana se encuentra sujeta a las fluctuaciones debidas a la energ\'ia t\'ermica del solvente. Si suponemos que la bicapa se 
encuentra en equilibrio con la soluci\'on acuosa, entonces, su energ\'ia est\'a caracterizada por las deformaciones que la membrana 
puede sufrir: a) la compresi\'on o expansi\'on de la bicapa en la direcci\'on lateral y b) las  deformaciones en la direcci\'on 
normal a la superficie (esto es, por la curvatura local de la superficie) \cite{safran}. Bajo estas condiciones, la energ\'ia libre 
(por mol\'ecula) asociada a la membrana, puede ser expresada como una funci\'on del \'area promedio de cada mol\'ecula, $\Sigma$,  
y de las curvaturas locales asociadas a la superficie. 
	
Los conceptos matem\'aticos necesarios para llegar a la energ\'ia el\'astica de las membranas pertenecen a la Geometr\'ia 
Diferencial de Superficies, por lo que conviene repasar algunos conceptos b\'asicos de dicho campo de la matem\'atica. 

\subsection{\label{sec:conceptos}Conceptos matem\'aticos previos}

\begin{figure*}
 \centering
 \includegraphics[width=0.70\textwidth]{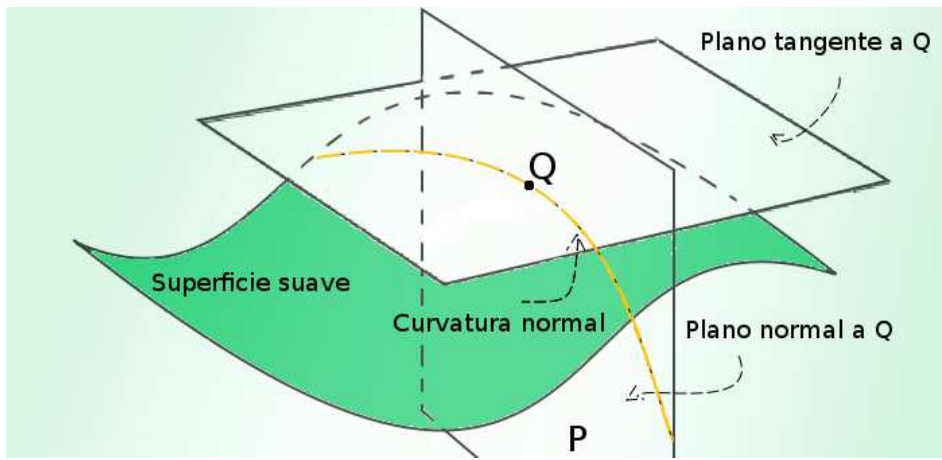}
  \caption{{\small Ilustraci\'on de la curvatura normal al punto $Q$ de una superficie.}}
 \label{fig:cnormales}
\end{figure*}

Consid\'erese un plano $P$, perpendicular al plano tangente (en un punto cualquiera) a la superficie que representa a la membrana.  
A la intersecci\'on de la superficie con el plano $P$ se le llama una ``curva normal''. Esta curva normal puede ser una l\'inea 
recta, sin curvatura, o puede poseer cierta curvatura, a la cual se le llama ``curvatura normal'', $c$. Estos conceptos se ilustran 
en la Figura \ref{fig:cnormales}. Para asociar un valor a la curvatura normal, notemos que una l\'inea recta (por definici\'on sin 
curvatura), puede verse como la circunferencia de un c\'irculo de radio infinito. Esto sugiere definir a la curvatura como el 
inverso del radio de curvatura: $c=1/R$. Como es de esperarse, para una l\'inea recta $c\rightarrow0$ puesto que $R\rightarrow\infty$. 
A la curvatura as\'i definida se le asocia un signo para distinguir las dos posibles orientaciones de la curva: 
concavidad hacia un lado o concavidad hacia el otro, direcciones a las cuales se asigna arbitrariamente el signo positivo o negativo. 

Notemos que, dado que el plano $P$ se defini\'o simplemente como perpendicular al plano tangente, existe una infinidad de posibles 
orientaciones para $P$. Dependiendo de dicha orientaci\'on, las diferentes curvas normales posibles tienen diferentes curvaturas. 
Sin embargo, puede demostrarse que hay dos direcciones particulares, perpendiculares entre s\'i, para las cuales la curvatura 
normal es un extremo (en un caso m\'aximo y en el otro m\'inimo). Dichas curvaturas normales extremas son llamadas las ``curvaturas 
principales'' de la superficie y se denotan mediante los s\'imbolos $c_{1}$ y $c_{2}$. Todas las dem\'as curvaturas normales en el 
mismo punto (as\'i como la curvatura local de cualquier curva contenida en la superficie y que pase por dicho punto) pueden ser 
expresadas en t\'erminos de las dos curvaturas principales. En otras pala\-bras, todas las deformaciones locales de la superficie 
en la direcci\'on normal al plano tangente pueden ser caracterizadas a primer orden por las curvaturas principales $c_{1}$ y $c_{2}$. 
Estos son los dos par\'ametros necesarios para caracterizar la curvatura de una membrana fluctuante.

\begin{figure}
 \centering
 \includegraphics[width=0.45\textwidth]{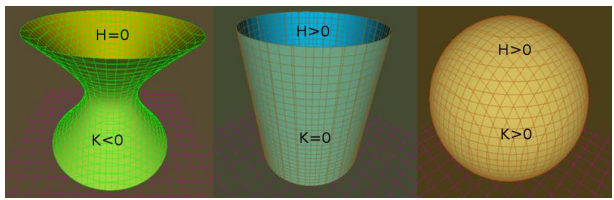}
 \caption{{\small Superficies con distinta curvatura media y gaussiana.}}
 \label{fig:cgaussiana}
\end{figure}

Sin embargo, es m\'as conveniente matem\'aticamente realizar un cambio de varia\-ble, de $c_{1}$ y $c_{2}$, a dos combinaciones de 
ellas, la llamada ``curvatura media'', $H = (c_{1} + c_{2})/2$, y la llamada ``curvatura gaussiana'', $K = c_{1} c_{2}$. $H$ es el 
par\'ametro mencionado al comentar el modelo de Canham en la introducci\'on (ecuaci\'on 1). De hecho, a segundo orden en las 
curvaturas principales, $\arrowvert H\arrowvert$ y $K$ son las \'unicas cantidades invariantes ante cualquier parametrizaci\'on 
de la superficie. Es importante se\~nalar que, dado que tanto $c_{1}$ y $c_{2}$ tienen asociado un signo, el valor de $H$ depende 
de la orientaci\'on elegida como positiva para la curvatura (cambiar la convenci\'on de signos, cambia el signo de $H$). El signo 
de $K$, por su parte, es independiente de esta elecci\'on. La Figura \ref{fig:cgaussiana} muestra algunas superficies con valores 
diferentes de las curvaturas media y gaussiana. 

\subsection{\label{sec:derivacion}Derivaci\'on de la energ\'ia de Helfrich}

Apliquemos ahora los conceptos geom\'etricos antes mencionados a la descripci\'on f\'isica de las bicapas lip\'i\-dicas. La 
energ\'ia libre de la bicapa (por unidad de \'area), $f$, debe ser invariante bajo cualquier parametrizaci\'on de la superficie, 
dado que es un concepto f\'isico que no depende de la descripci\'on matem\'atica utilizada. Luego entonces, dicha energ\'ia debe 
ser una funci\'on de $H$ y $K$. 

En el equilibrio las bicapas pueden considerarse aproximadamente planas, al menos localmente; las deformaciones respecto a dicho 
estado est\'an caracterizadas por curvaturas muy peque\~nas; es decir, el radio de curvatura correspondiente es mucho mayor que 
el espesor de la bicapa. Esto permite desarrollar la funci\'on $f=f(H,K)$ mediante una serie de Taylor (a segundo orden en las 
curvaturas $c_{1}$ y $c_{2}$):
\begin{equation}
	f(H,K)\approx f_{0}+f_{1}H+f_{2}H^{2}+f_{3}K
	\label{eq:etaylor}
\end{equation}
donde las constantes $f_{0}$, $f_{1}$, $f_{2}$ y $f_{3}$ dependen en principio del \'area promedio por mol\'ecula; en particular 
$f_{0}$ corresponde a la energ\'ia libre de una bicapa plana. 

N\'otese que el t\'ermino lineal en $H$ depende de la orientaci\'on de la superficie. Cuando las monocapas que forman la bicapa 
tienen la misma composici\'on, no hay raz\'on alguna para dicha dependencia en la orientaci\'on y, en tal caso, la constante 
$f_{1}$ debe ser cero. En general, sin embargo, existe la posibilidad de una asimetr\'ia entre las monocapas y es precisamente 
el t\'ermino lineal en $H$ el que toma en cuenta este efecto.

Completando el binomio cuadrado respecto a $H$, la ecuaci\'on (\ref{eq:etaylor}) puede reescribirse como:
\[
f=f_{2}\left(H+\dfrac{f_{1}}{2f_{2}}\right)^{2}+\left(f_{0}-\dfrac{f_{1}^{2}}{4f_{2}}\right)+f_{3}K.
\]
Si entonces definimos las cantidades:
\begin{eqnarray}
  \sigma\equiv f_{0} - \dfrac{f_{1}^{2}}{4 f_{2}},\nonumber\\
	\kappa\equiv 2f_{2},\hspace{0.9cm}\nonumber\\
	c_{0}\equiv -\dfrac{f_{1}}{2f_{2}} \hspace{0.5cm}\mbox{ y }\nonumber\\
	\bar{\kappa}\equiv f_{3}\hspace{1.2cm}\nonumber
\end{eqnarray}
	la energ\'ia libre, por unidad de \'area, de la bicapa toma la forma
\begin{equation}
	f=\dfrac{1}{2}\kappa\left(H-c_{0}\right)^{2}+\bar{\kappa}K + \sigma,
	\label{eq:helfrich}
\end{equation}
donde $\sigma$, $\kappa$, $\bar{\kappa}$ y $c_{0}$ son la tensi\'on superficial, el m\'odulo de curvatura media, el m\'odulo de 
curvatura gaussiana y la curvatura espont\'anea de la membrana, respectivamente. Colectivamente esas variables son conocidas como 
los ``par\'ametros de Helfrich''. Dichos par\'ametros son independientes de la forma geom\'etrica de la membrana pero dependen en 
principio de su composici\'on qu\'imica. 

La energ\'ia total de la bicapa (conocida como la energ\'ia de Helfrich) puede ser calculada integrando sobre el \'area total de 
la misma:
\begin{equation}
	F=\int{\left[\dfrac{1}{2}\kappa\left(H-c_{0}\right)^{2}+\bar{\kappa}K + \sigma \right]}dA.
	\label{eq:helfricht}
\end{equation}

Wolfgang Helfrich lleg\'o a estas ecuaciones en 1973 en un af\'an de describir la elasticidad de las membranas biol\'ogicas 
\cite{helfrich}. La derivaci\'on original publicada ese a\~no, se basa en una analog\'ia con la energ\'ia libre para cristales 
l\'iquidos propuesta por Frank \cite{frank}. Sin embargo, la forma en que aqu\'i arribamos a tal relaci\'on fue desarrollada por 
Safran en 1994. Hemos elegido esta \'ultima derivaci\'on debido a la sencillez y elegancia. 

Es importante notar que si no hay fuerzas externas que tensen la membrana, el \'area de la bicapa puede considerarse constante. 
Esto es as\'i porque las colas hidrof\'obicas de las mol\'eculas de fosfol\'ipido son casi totalmente insolubles en agua y est\'an 
agrupadas de manera que el \'area por mol\'ecula se mantiene pr\'acticamente constante. Modificar dicha \'area requiere de 
energ\'ias mayores a la necesaria para simplemente reacomodar las mol\'eculas de la bicapa de modo que \'esta se curve. Por esta 
raz\'on, en ausencia de fuerzas externas, el t\'ermino de la energ\'ia de Helfrich que contiene la tensi\'on superficial, $\sigma$, 
es una constante sin relevancia al momento de minimizar la energ\'ia respecto a las posibles deformaciones de la superficie. En la 
discusi\'on subsiguiente ignoraremos este t\'ermino, excepto cuando se mencione expl\'icitamente lo contrario.

 \subsection{\label{sec:significado}Significado f\'isico de los par\'ametros en la energ\'ia de Helfrich}

El significado de los par\'ametros de Helfrich puede entenderse sencillamente anali\-z\'andolos por separado. 

\begin{figure}
 \centering
 \includegraphics[width=0.40\textwidth]{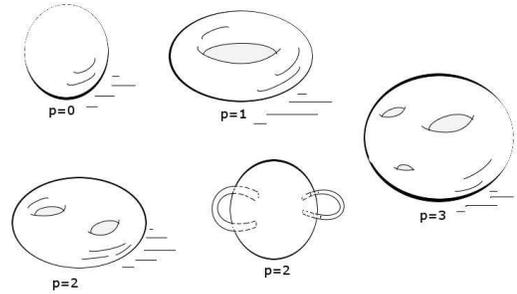}
 \caption{{\small Superficies con distinto g\'enero.}}
 \label{fig:asideros}
\end{figure}

Para comentar la f\'isica que se desprende del m\'odulo de curvatura media, $\kappa$, es conveniente empezar haciendo algunas 
observaciones acerca del t\'ermino de la curvatura gaussiana, el segundo de la ecuaci\'on (\ref{eq:helfrich}). El llamado teorema 
de Gauss-Bonnet \cite{kreiszing} muestra que este t\'ermino es una constante si la topolog\'ia de la membrana no cambia. En efecto, 
este teorema estipula que para una superficie cerrada de topolog\'ia fija, la integral de la curvatura gaussiana es una constante,
\begin{equation}
	\int{K}dA=4\pi\left(1-p\right),
\label{eq:gbonet}
\end{equation}
donde $p$ denota el g\'enero de la superficie, concepto topol\'ogico que est\'a relacionado con la cantidad de poros o perforaciones 
que tiene la superficie. Algunos ejemplos de superficies con distinto g\'enero se muestran en la Figura \ref{fig:asideros}. La 
conclusi\'on principal del teorema de Gauss-Bonnet es que, al estudiar la estructura de m\'inima energ\'ia de membranas de 
topolog\'ia fija, el t\'ermino que involucra la curvatura gaussiana puede ser ignorado, pues es constante mientras la topolog\'ia 
no cambie. En otras palabras, la energ\'ia de curvatura asociada al m\'odulo gaussiano no cambia mientras la topolog\'ia de la 
membrana sea la misma. 

Bajo las condiciones anteriores, la energ\'ia el\'astica de Helfrich toma la forma de un potencial arm\'onico 
$F=\dfrac{1}{2}\kappa\int{\left(H-c_{0}\right)^{2}}dA$ que depende de la curvatura media $H$ en forma an\'aloga a como la 
energ\'ia potencial de un resorte depende de su deformaci\'on. Entonces, el m\'odulo de curvatura media, $\kappa$, es an\'alogo a 
la constante de r\'igidez de un resorte, y puede interpretarse como una constante de r\'igidez (ante deformaciones de curvatura) 
de la membrana. Cuando $\kappa$  tiene valores peque\~nos, la membrana es flexible; cuando $\kappa$ asume valores grandes, la 
membrana es r\'igida. Entonces, el t\'ermino de curvatura media est\'a relacionado con las fluctuaciones de forma u ondulaciones 
(sin cambio topol\'ogico) que experimenta la membrana por efectos de la energ\'ia t\'ermica del solvente.

Por su parte, la curvatura espont\'anea $c_{0}$ es la curvatura media que adquiere la bicapa cuando est\'a libre de tensi\'on, 
jugando un papel an\'alogo al de la longitud de equilibrio en el caso del resorte. Hicimos notar anteriormente que para monocapas 
id\'enticas la constante $f_{1}$ de la ecuaci\'on (\ref{eq:etaylor}) debe ser cero y, por tanto, $c_{0}$ debe ser nula tambi\'en. 
Es claro entonces que la curvatura espont\'anea est\'a asociada a la asimetr\'ia de las monocapas que forman la bicapa. Esta 
curvatura es debida al hecho de que en general las monocapas se forman por distintas  mol\'eculas anfif\'ilicas ocasionando que 
las bicapas tiendan a curvarse espont\'aneamente hacia alguno de sus lados cuando dos de estas monocapas se superponen. Por supuesto 
que cuando la bicapa est\'a formada por un solo tipo de mol\'eculas, no existe curvatura espont\'anea y el estado de equilibrio 
tiende a ser el de una membrana plana. En resumen podemos decir que la ecuaci\'on (\ref{eq:helfrich}) establece que cuando no hay 
cambios de topolog\'ia en el sistema la curvatura media que m\'inimiza la energ\'ia libre corresponde a la curvatura espont\'anea. 
El radio de equilibrio de la membrana es $R_{0}=1/c_{0}$. 

Finalmente, para tener una idea del significado f\'isico del m\'odulo de curvatura gaussiana, $\bar{\kappa}$, imaginemos una bicapa 
formada por monocapas id\'enticas ($c_{0} = 0$) que puede cambiar de topolog\'ia. El t\'ermino de la curvatura media, 
$F=\dfrac{1}{2}\kappa\int{H^{2}}dA$, es m\'inimo para una superficie con $H = 0$, que puede ser un plano pero tambi\'en una 
superficie localmente de tipo ``silla de montar'' (Ver figura \ref{fig:cgaussiana}). 

Puede ahora notarse que el t\'ermino de la curvatura gaussiana, $\int{\bar{\kappa}K}dA$, favorece superficies con distintas 
topolog\'ias, dependiendo del valor del m\'odulo el\'astico gaussiano. En efecto, cuando $\bar{\kappa}\geq 0$, la energ\'ia 
relacionada con el t\'ermino gaussiano se minimiza si los radios de las curvaturas principales tienen diferente signo, es decir, 
si la curvatura gaussiana es negativa, como en el caso de superficies tipo ``silla de montar'', prevalecientes en las fases esponja 
y c\'ubicas. Por otra parte, cuando $\bar{\kappa}\leq 0$, el t\'ermino gaussiano favorece superficies de curvatura gaussiana 
positiva (radios de curvatura en la misma direcci\'on), contribuyendo a la estabilidad de agregados cerrados tales como ves\'iculas.

Esta breve discusi\'on nos permite vislumbrar el tipo de fen\'omenos f\'isicos que afectan a una membrana y que pueden ser descritos 
con la ecuaci\'on (\ref{eq:helfrich}): forma de equilibrio, fluctuaciones t\'ermicas y cambios de topolog\'ia. Por esta raz\'on, la 
energ\'ia de Helfrich ha sido utilizada en una gran cantidad de trabajos que describen algunas propiedads f\'isicas de las membranas. 
De hecho, la posibilidad de calcular la energ\'ia  libre de una bicapa ha permitido modelar sistemas biol\'ogicos reales  como el 
aparato de Golgi \cite{golgi} o la forma discoidal de los gl\'obulos rojos \cite{deuling1, fung}. 
La forma en que se usa la relaci\'on (\ref{eq:helfrich}) en los distintos modelos consiste en sumar las diferentes contribuciones 
energ\'eticas para despu\'es minimizar la energ\'ia total a fin de obtener informaci\'on sobre la forma o las fluctuaciones del 
sistema. En la siguiente secci\'on veremos una forma matem\'aticamente conveniente de expresar las fluctuaciones de una membrana.
lo de curvatura media $\kappa$ de una membrana. En los diferentes m\'etodos se aprovecha 
el hecho de que valores diferentes de $\kappa$ hacen que la membrana sea m\'as o menos flexible y por lo tanto tenga diferente 
susceptibilidad a las fluctuaciones t\'ermicas del solvente. As\'i, una membrana muy flexible fluct\'ua m\'as que una membrana 
r\'igida, lo cual ocasiona consecuencias medibles en varios fen\'omenos f\'isicos. Por ejemplo, la dispersi\'on de radiaci\'on (luz, 
rayos X o neutrones) es diferente si la membrana es flexible o si es r\'igida. De igual forma, cuando una ves\'icula gigante es 
aspirada con una micropipeta, el cambio aparente en su \'area es diferente, dependiendo de si la membrana que la forma es flexible 
o no. Analizaremos este \'ultimo experimento con m\'as detalle en la siguiente secci\'on para mostrar una manera de medir $\kappa$. 
Para ello, primeramente derivamos algunas expresiones matem\'aticas para las fluctuaciones de una membrana.

\subsection{\label{sec:monge}Energia de Helfrich en la parametrizaci\'on de Monge}

\begin{figure}
 \centering
 \includegraphics[width=0.30\textwidth]{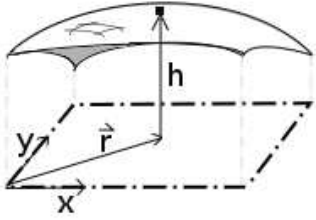}
 \caption{{\small Representaci\'on esquem\'atica de la parametrizaci\'on de una membrana con peque\~nas fluctuaciones t\'ermicas.}}
 \label{fig:pmonge}
\end{figure}

Por simplicidad, vamos a analizar el caso de una membrana que en promedio se encuentra horizontal respecto a un plano de 
referencia pero que presenta peque\~nas fluctuaciones t\'ermicas en la direcci\'on perpendicular al plano. La posici\'on sobre el 
plano de referencia se describe con un vector de posici\'on bidimensional $\vec{r} = (x,y)$. A un tiempo dado, la conformaci\'on de 
la bicapa puede caracterizarse mediante la altura en cada punto $z = h(\vec{r})$ (Ver Figura \ref{fig:pmonge}). A la descripci\'on 
matem\'atica de la membrana realizada as\'i, se le llama ``parametrizaci\'on de Monge'' y es una de las maneras m\'as simples de 
parametrizar una superficie. Supondremos que la membrana tiene dimensiones $L\times L$ en el plano de referencia. Adem\'as, por 
simplicidad, consideremos una bicapa con dos monocapas id\'enticas de modo que $c_{0} = 0$. 

Lo primero que veremos es la forma que adopta la curvatura media en la repre\-sentaci\'on de Monge. Comencemos por analizar 
un caso m\'as simple: una curva unidimensional en un espacio bidimensional. Dicha curva puede pensarse como la trayectoria de 
un m\'ovil, $\vec{R}(x)$, donde el par\'ametro $x$ puede pensarse moment\'anemente como un tiempo. La ``velocidad'' del 
m\'ovil ser\'ia $\vec{v}=\vec{R}'(x)$ y su ``aceleraci\'on'', $\vec{a}=\vec{R}''(x)$. El ap\'ostrofe denota derivada con 
respecto a $x$. La aceleraci\'on centr\'ipeta es igual al cuadrado de la velocidad por la curvatura en el punto en cuesti\'on, 
$a_{c}=\frac{v^{2}}{r_{c}}=\frac{\| \vec{R}'(x) \|^{2}}{r_{c}}$. Por tanto, para calcular la curvatura necesitamos calcular 
de manera independiente la aceleraci\'on centr\'ipeta. Esto lo haremos aprovechando que, a su vez, $a_{c}$ es igual a la 
componente de la aceleraci\'on perpendicular a la velocidad. Esta componente puede ser obtenida recordando que el \'area $A$ 
del paralelogramo formado por los vectores $\vec{R'}$ y $\vec{R''}$ es igual al  producto de su base, ${\| \vec{R}'(x) \|}$ 
por su altura, $a_{c}$, de donde $a_{c}=\frac{A}{\| \vec{R}'(x) \|}$. Ahora recordemos que el \'area (orientada) del 
paralelogramo formado por los vectores $\vec{R}'(x)$ y $\vec{R}''(x)$ es igual al determinante 
$\det\left[\vec{R}'(x), \vec{R}''(x) \right]$ (el signo del determinante est\'a relacionado a la orientaci\'on de un vector 
respecto al otro). Por lo tanto, la aceleraci\'on centr\'ipeta est\'a dada por 
$a_{c}=\frac{\det\left[ \vec{R}'(x), \vec{R}''(x) \right] }{\| \vec{R}'(x) \|}$. Despejando, la curvatura en dicho punto 
es entonces: $c = \frac{1}{r_{c}} = \frac{\det\left[ \vec{R}'(x), \vec{R}''(x) \right]}{\| \vec{R}'(x) \|^{3}}$. 
Esta f\'ormula es general para cualquier curva. 

Para hacer la analog\'ia con el caso que queremos estudiar podemos pensar en una curva que es casi recta, excepto por peque\~nas 
fluctuaciones, y que por lo tanto puede representarse por el an\'alogo de la parametrizaci\'on de Monge, $\vec{R}(x) = (x, h(x))$. 
Es ahora claro que, a primer orden en $h$ y sus derivadas, la curvatura es: 
$c = \frac{\det\left[\left(1, h_{x}\right), \left(0, h_{xx}\right)\right]}{\left( 1+ h_{x}^{2} \right)^{3/2}} \approx h_{xx}$.

An\'alogamente, en el caso tridimensional que nos interesa, la curvatura media de una superficie est\'a dada, a segundo orden en 
$h$ y sus derivadas, por la siguiente f\'ormula:
\[
	H\approx\nabla^{2} h 
\]
donde $\nabla^{2}$ representa el operador laplaciano, el cual corresponde a la segunda derivada del p\'arrafo anterior. Por su parte, 
el elemento diferencial de \'area est\'a dado por 
$\sqrt{1+ h_{x}^{2}+ h_{y}^{2}} dx dy \approx \left[ 1 + \frac{1}{2} \left( \nabla h \right)^{2} \right] dx dy$. Sustituyendo estas 
expresiones en la ecuaci\'on (\ref{eq:helfricht}), tenemos que la energ\'ia de Helfrich para membranas d\'ebilmente 
curvadas puede escribirse en forma aproximada como:
\begin{equation} 
	F=\dfrac{1}{2}\int{\left\lbrace \kappa\left(\nabla^{2} h\right)^{2}+\sigma\left(\nabla h\right)^{2}\right\rbrace }dxdy,
\label{eq:emonge}
\end{equation}
  donde estamos ignorando el t\'ermino constante. Esta ecuaci\'on supone adem\'as que las fluctuaciones t\'ermicas afectan la forma 
de la membrana pero no provocan cambios en su topolog\'ia, por lo cual se ha ignorado el t\'ermino gaussiano, que como vimos permanece 
constante mientras no cambie la topolog\'ia. Esta manera de escribir la energ\'ia el\'astica de Helfrich es matem\'aticamente 
conveniente para describir las fluctuaciones de forma en la membrana, lo cual ha llevado a concebir m\'etodos experimentales para medir 
la constante el\'astica $\kappa$.

\subsection{\label{sec:forma}Fluctuaciones de forma de una membrana}

Para estudiar las fluctuaciones de una membrana, vamos a considerar sus diversos modos de deformaci\'on. Para ello, haremos una 
expansi\'on en serie de Fourier de la funci\'on que describe la forma de la membrana en la parametrizaci\'on de Monge, $h(\vec{r})$. 
Esta expansi\'on queda de la siguiente forma:
\begin{equation}
 h\left(\vec{r}\right)=\sum_{\vec{q}}{h_{\vec{q}}\cdot\exp{\left\lbrace i\vec{q}\cdot\vec{r}\right \rbrace}}
\label{eq:fourier}
\end{equation}
donde $\vec{q}=\frac{2\pi}{L}\left(n_{x},n_{y}\right)$ y $n_{x}$, $n_{y}$ son n\'umeros enteros. Los coeficientes $h_{\vec{q}}$ 
pueden verse como ``coordenadas'' en el espacio de Fourier y representan la contribuci\'on a las fluctuaciones de cada modo de 
deformaci\'on.

Para utilizar esta expansi\'on en la expresi\'on de la energ\'ia de Helfrich (ecuaci\'on \ref{eq:emonge}), derivamos la expresi\'on 
anterior para obtener el gradiente y el laplaciano de $h(\vec{r})$:  
\begin{eqnarray}
& &\left(\nabla h\right)^{2}=-\sum_{\vec{q}_{1},\vec{q}_{2}}{\left(\vec{q}_{1}\cdot\vec{q}_{2}\right)h_{\vec{q}_{1}} h_{\vec{q}_{2}}
\exp{\left\lbrace i(\vec{q}_{1}+\vec{q}_{2})\cdot \vec{r} \right\rbrace}},\vspace{0.5cm}\\
& &\left(\nabla^{2} h\right)^{2}=\sum_{\vec{q}_{1},\vec{q}_{2}}{q_{1}^{2}q_{2}^{2}h_{\vec{q}_{1}} h_{\vec{q}_{2}} 
\exp{\left\lbrace i(\vec{q}_{1}+\vec{q}_{2})\cdot\vec{r}\right\rbrace}}.
\end{eqnarray}
Con estas ecuaciones, la energ\'ia libre de Helfrich, (\ref{eq:emonge}), puede reescribirse como
\begin{widetext}
\begin{equation}
 F=\dfrac{1}{2}\int_{L\times L}{\sum_{\vec{q}_{1},\vec{q}_{2}}h_{\vec{q}_{1}}   h_{\vec{q}_{2}} \left[ \kappa q_{1}^{2} q_{2}^{2}  - \sigma 
\left( \vec{q}_{1} \cdot \vec{q}_{2} \right) \right]\exp{\left\lbrace i(\vec{q}_{1}+\vec{q}_{2})\cdot\vec{r}\right\rbrace }}d^{2} \vec{r}.
\end{equation}
\end{widetext}
Esta \'ultima expresi\'on puede escribirse en una forma m\'as simple intercambiando la integral con la sumatoria y usando la 
representaci\'on de Fourier de la delta de Kronecker:
\begin{equation}
\int_{L\times L}{\exp{\left\lbrace -i\vec{q}\cdot\vec{v}\right\rbrace}}d^{2}\vec{v}=L^{2}\delta_{\vec{q},0}.
\label{eq:kronecker}
\end{equation}
 De aqu\'i, $F=\dfrac{L^{2}}{2}\sum_{\vec{q}}{h_{\vec{q}} h_{-\vec{q}}\left(\kappa q^{4}+\sigma q^{2}\right)}$.  Por otra parte, 
el hecho de que $h(\vec{r})$ sea real y la ecuaci\'on (\ref{eq:fourier}) nos llevan a que los modos complejos de Fourier son tales 
que $h_{-\vec{q}}=h^{*}_{\vec{q}}$. Todo esto finalmente nos lleva a que la energ\'ia de Helfrich es:
\begin{equation}
 	F=\dfrac{L^{2}}{2}\sum_{\vec{q}}{| h_{\vec{q}} |^{2} \left(\kappa q^{4}+\sigma q^{2}\right)}.
\label{eq:energiaq}
\end{equation}

Observemos ahora que la energ\'ia obtenida es arm\'onica en estas nuevas ``coordenadas'' $| h_{\vec{q}} |$. Si ahora se aplica el 
teorema de equipartici\'on de la energ\'ia (Ver Ap\'endice \ref{sec:equiparticion}), las amplitudes cuadr\'aticas medias de los 
diferentes modos de oscilaci\'on en la membrana est\'an dadas por:

\begin{equation}
 	\left\langle h^{2}_{\vec{q}}\right\rangle=\dfrac{K_{B}T}{L^{2}\left(\kappa q^{4}+\sigma q^{2}\right)}. 
 	\label{eq:equiparticion}
\end{equation}

Com\'unmente a esta funci\'on se le da el nombre de espectro de fluctuaciones o factor de estructura. Vemos en esta ecuaci\'on que, 
como era de esperarse, la amplitud de las fluctuaciones de la membrana aumenta al aumentar la temperatura. Igualmente, dado que la 
amplitud de cada modo es inversamente proporcional al m\'odulo el\'astico de curvatura media, $\kappa$, vemos que las fluctuaciones 
son m\'as grandes mientras m\'as peque\~no sea este m\'odulo, y vice\-versa.
 
Por otra parte, dos modos diferentes est\'an desacoplados y, por tanto (Ver Ap\'en\-dice \ref{sec:equiparticion}), 
\begin{equation}
 	\left\langle h_{\vec{q}_{1}} h_{\vec{q}_{2}}\right\rangle = 0,
\label{eq:desacopl}
\end{equation}
si $\vec{q}_{1} \neq -\vec{q}_{2}$.
Utilizando este hecho es posible ver que la amplitud total de la fluctuaci\'on de la membrana en el espacio real es simplemente la 
suma sobre todos los modos individuales, es decir, 
\[
\left\langle h^{2}\right\rangle=\sum_{q}{\left\langle h^{2}_{\vec{q}}\right\rangle}.
\label{eq:suma-aplituda}
\]
Esta relaci\'on es muy importante pues su evaluaci\'on permite tener una idea sobre los par\'ametros de los que depende la amplitud 
de las fluctuaciones de forma en la membrana. Para calcular la amplitud total en forma aproximada la suma se puede transformar en 
una integral:
\[
\sum_{q}\rightarrow 2\pi\int_{q_{min}}^{q_{max}}{\left( \dfrac{L}{2\pi}\right)^{2}}qdq.
\] 
Cabe hacer notar que en esta ecuaci\'on, los l\'imites de integraci\'on se deben elegir de tal manera que se sumen las fluctuaciones 
que tienen sentido f\'isico. Por ejemplo, no tiene sentido considerar fluctuaciones cuya longitud de onda sea mayor que la dimensi\'on 
mayor de la membrana ($L$). De igual forma, no tiene sentido considerar fluctuaciones m\'as peque\~nas que la escala molecular. Para 
esta \'ultima escala espacial, como distancia de corte se puede tomar una distancia del orden del espesor de la membrana, pues este 
tama\~no est\'a determinado por la longitud de las mol\'eculas anfif\'ilicas. 
Con estos razonamientos, con una buena aproximaci\'on se pueden proponer l\'imites de integraci\'on dados por: 
$q_{max}=2\pi/a$ y $q_{min}=2\pi/L$, donde $a$ es del orden del grosor de la membrana y $L$ es la longitud total de \'esta. 

Tomando en cuenta las consideraciones expuestas,
\[
 \left\langle h^{2}\right\rangle=2\pi\left( \dfrac{L}{2\pi}\right)^{2}\int_{q_{min}}^{q_{max}}{\dfrac{K_{B}T}{L^{2}
\left(\kappa q^{4}+\sigma q^{2}\right)}q}dq,
\]
	es decir,
\begin{equation}
\left\langle h^{2}\right\rangle=\dfrac{K_{B}T}{4\pi\sigma}\ln\dfrac{q^{2}_{max}\left( q^{2}_{min}\kappa+\sigma\right)}
{q^{2}_{min}\left( q^{2}_{max}\kappa+\sigma\right)}.
\label{eq:amplitud}
\end{equation}

En particular, cuando $q^{2}_{min}\ll q^{2}_{max}$ y $\sigma\rightarrow 0$, la amplitud total toma la forma:
\begin{equation}
	\left\langle h^{2}\right\rangle\approx\dfrac{K_{B}T}{16\pi^{3}\kappa}L^{2}.
\label{eq:aplituda}
\end{equation}

Vemos entonces que el cuadrado de la amplitud total de las fluctuaciones es (en promedio) proporcional a la temperatura e 
inversamente proporcional al m\'odulo el\'astico de curvatura media, $\kappa$, lo cual corrobora la idea intuitiva de que 
mientras m\'as flexible es la membrana, mayores son sus fluctuaciones.

\subsection{\label{sec:experimental}Medici\'on experimental del m\'odulo de curvatura media de una ves\'icula mediante 
manipulaci\'on con micropipetas}

Hemos dicho que existen diferentes m\'etodos experimentales que permiten medir el m\'odulo el\'astico de curvatura media, 
$\kappa$. En esta secci\'on comentaremos uno de ellos, basado en la t\'ecnica de aspiraci\'on de ves\'iculas con micropipetas.

Este m\'etodo aprovecha que las ves\'iculas gigantes, preparadas con el m\'etodo de electroformaci\'on, son f\'acil\-mente observables 
con ayuda de un microscopio \'optico \cite{evans1, rawicz}. En general, las ves\'iculas tienen una aparien\-cia de esferas perfectas. 
Sin embargo, debido a que la membrana que las conforma es flexible, su superficie presenta una cierta corrugaci\'on cambiante con 
el tiempo debido a las fluctuaciones t\'ermicas \cite{helfrich2}. Esas peque\~nas desviaciones respecto de la esfera se aprovechan 
para la medici\'on experimental del m\'odulo de curvatura media $\kappa$ de la membrana. 

Para ello se utiliza la t\'ecnica de micromanipulaci\'on o microaspiraci\'on con micropipetas. En este m\'etodo, ves\'iculas 
gigantes preparadas con el m\'etodo de electroformaci\'on (tama\~no del orden de 20 micras) se aspiran con un capilar muy delgado 
de vidrio (di\'ametro del orden de 5 micras). Mediante el capilar se aplica una presi\'on de succi\'on a la ves\'icula (Figura 
\ref{fig:micropipetas}). Como la ves\'icula es deformable, parte de ella penetra en la secci\'on cil\'indrica de la micropipeta. 
El \'area de la ves\'icula cambia y su forma ya no es exactamente esf\'erica, sino que se debe considerar el trozo de membrana 
dentro del capilar. Evidentemente la distancia de penetraci\'on depende tanto de la presi\'on aplicada como de la elasticidad de 
la membrana. As\'i, evaluando el cambio del \'area de la ves\'icula (a partir de fotograf\'ias), y conociendo la presi\'on aplicada, 
puede determinarse el m\'odulo de curvatura media de las membranas, $\kappa$. 

Tanto el cambio de \'area de la membrana, como la tensi\'on de la ves\'icula pueden escribirse en t\'erminos de par\'ametros 
controlables o medibles en el experimento. 

Por ejemplo, utilizando argumentos de equilibrio termodin\'amico, Henriksen e Ipsen \cite{ipsen} mostraron que la tensi\'on de la 
membrana, $\sigma$, est\'a relacionada a la presi\'on de succi\'on, $\Delta p$, mediante la f\'ormula: 
\begin{figure*}
	\centering
	\includegraphics[width=0.70\textwidth]{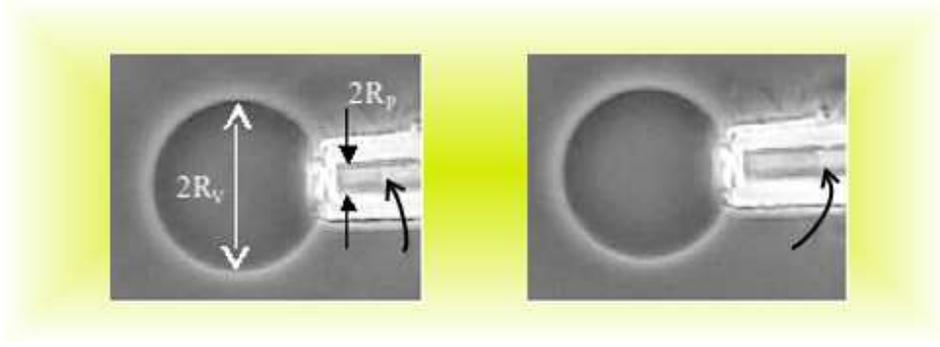}
	\caption{{\small Fotograf\'ias tomadas en un microscopio \'optico de una ves\'icula de SOPC aspirada dentro de una 
micropipeta. La presi\'on de aspiraci\'on es mayor en la foto de la derecha. Esto se refleja en la longitud diferente de la secci\'on
 de membrana dentro de la pipeta (flechas oblicuas). Fotos cortes\'ia de Gerardo Paredes, estudiante de doctorado en la 
Universidad de Sonora.}}
	\label{fig:micropipetas}
\end{figure*}
\begin{equation}	
 \sigma = \dfrac{\Delta p R_p}{2 \left( 1 - \dfrac{R_p}{R_v} \right)},
\end{equation}
donde $R_p$ es el radio de la pipeta y $R_v$ el radio de la ves\'icula. 

\begin{figure}
 \centering
 \includegraphics[width=0.45\textwidth]{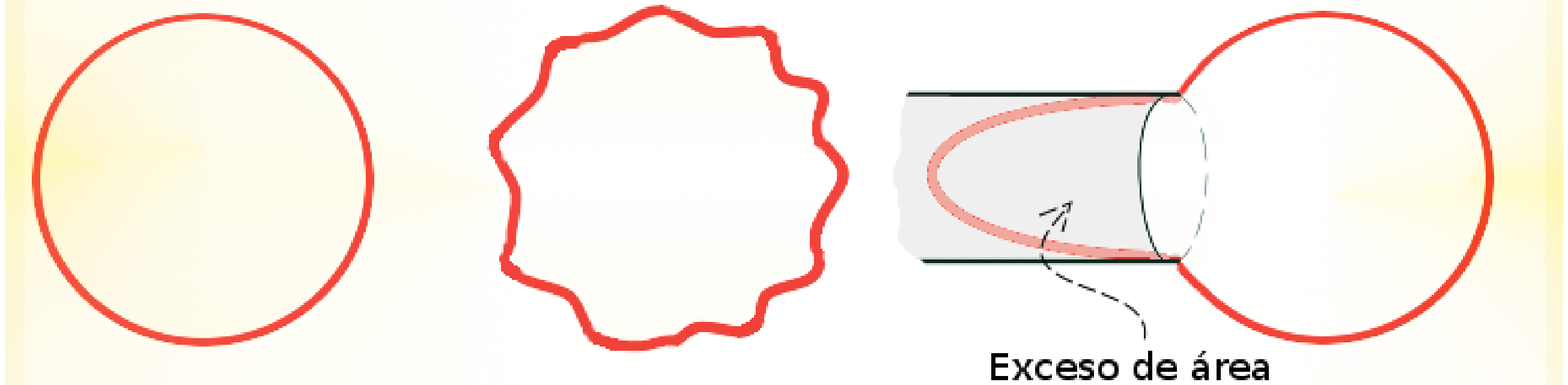}
  \caption{{\small Representaci\'on esquem\'atica de una membrana sin fluctuaciones t\'ermicas y de una membrana con fluctuaciones. 
Al aspirar con una micropipeta la ves\'icula fluctuante, las fluctuaciones se suprimen. El exceso aparente 
de \'area (porci\'on cil\'indrica de membrana dentro de la pipeta) es el \'area asociada a las fluctuaciones de forma antes de la 
aspiraci\'on.}}
  \label{fig:arrugada}
\end{figure}

Para evaluar el cambio del \'area de la ves\'icula al someterse a una presi\'on de succi\'on, recordemos que la membrana presenta 
fluctuaciones t\'ermicas que la deforman ligera pero constantemente. Una representaci\'on esquem\'atica de estas fluctuaciones 
en un momento dado se muestran en la Figura \ref{fig:arrugada}. En estas condiciones, el efecto de una presi\'on de succi\'on 
puede analizarse en dos reg\'imenes cualitativamente dife\-rentes, los cuales dependen de la magnitud de la presi\'on de 
succi\'on y, por lo tanto, de la tensi\'on $\sigma$:
\begin{enumerate}
 \item  Cuando la tensi\'on $\sigma$ a la cual est\'a sometida la membrana es peque\~na, el efecto de la presi\'on aplicada es reducir 
la amplitud de las fluctuaciones t\'ermicas, alisando la forma esf\'erica de la ves\'icula, ver Figura \ref{fig:arrugada}. Entonces, 
al penetrar la ves\'icula en el capilar, el \'area de la membrana permanece constante. El aparente incremento de \'area (cilindro de 
longitud $L$ en el capilar) viene de la supresi\'on de las fluctuaciones de curvatura alrededor de la forma esf\'erica (Figura 
\ref{fig:micropipetas}). Evidentemente que este efecto est\'a directamente relacionado con la flexibilidad o rigidez de la membrana. 
Si la presi\'on de succi\'on est\'a en este reg\'imen es posible medir el m\'odulo de curvatura media de la membrana, $\kappa$.

\item Cuando la tensi\'on $\sigma$ aplicada a la membrana es grande, esto trae como efecto, adem\'as de desaparecer las fluctuaciones 
de curvatura de la membrana, un estiramiento del \'area de la misma. En este caso, el \'area de la membrana se incrementa. Tambi\'en en 
este reg\'imen es posible medir lo que se conoce como la ``constante de compresibilidad de \'area'', $K_A$, que es el equivalente 
bidimensional a la compresibilidad volum\'etrica de un fluido \cite{hung}.
\end{enumerate}

Aqu\'i estamos interesados en mostrar c\'omo es posible medir el m\'odulo de curvatura media, $\kappa$. Por esta raz\'on 
concentraremos nuestra discusi\'on en el reg\'imen de peque\~nas tensiones. Vamos a analizar con detalle el caso m\'as sencillo de 
una membrana plana, aunque fluctuante, de dimensiones $L \times L$. El resultado final es id\'entico para una ves\'icula esf\'erica 
pero es necesario mayor trabajo matem\'atico para deducirlo.

 Primero consideremos que el \'area real de la membrana (incluyendo fluctuaciones t\'ermicas) no es exactamente el producto de sus 
dimensiones, $L\times L$. Esto es debido a que, por las fluctuaciones t\'ermicas, la membrana no es perfectamente plana, sino que en 
cada instante presenta cierta rugosidad. El \'area real puede calcularse integrando el elemento de \'area que describimos en la 
secci\'on anterior:
\begin{eqnarray*}
 &A&=\int_{L \times L}{\sqrt{1+\left(\nabla h(\vec{r}) \right)^{2}} }d^{2} \vec{r}\\
& &\approx \int_{L \times L}{\left(1+\frac{1}{2} 
\left[ \nabla h(\vec{r}) \right]^{2} \right) }d^{2} \vec{r},
\end{eqnarray*}
donde, como en la secci\'on anterior, $h \left( \vec{r} \right)$ es la altura de la membrana respecto al plano de referencia y su 
expansi\'on de Fourier est\'a dada por la ecuaci\'on (\ref{eq:fourier}). Obteniendo el gradiente encontramos que
\[
 \Delta A= -\dfrac{1}{2}\int_{L \times L}d^{2} \vec{r}{\sum_{\vec{q}_{1},\vec{q}_{2}}{\left( \vec{q}_{1} \cdot \vec{q}_{2} \right) 
h_{\vec{q}_{1}}h_{\vec{q}_{2}} e^{\left\{ i \left( \vec{q}_{1} + \vec{q}_{2}\right) \cdot \vec{r}\right\}}}} ,
\]
donde $\Delta A = A - A_{0}$ es la diferencia entre el \'area real de la membrana, $A$, y el \'area aparente al microscopio (es 
decir, el \'area de la membrana plana de referencia), $A_{0} = L^{2}$.  Utilizando la ecuaci\'on (\ref{eq:kronecker}) esta 
expresi\'on se simplifica: 
\[
 \Delta A=  \dfrac{L^{2}}{2} \sum_{\vec{q}}{q^{2} \| h_{\vec{q}} \|^{2}}.
\]
Podemos ahora tomar el promedio de $\Delta A$ y utilizar la ecuaci\'on (\ref{eq:equiparticion}) de la secci\'on anterior para obtener 
el si\-guiente resultado:
\[
 \left\langle \Delta A\right\rangle =\dfrac{1}{2}\sum_{\vec{q}}{\dfrac{K_{B}T}{q^{2}\kappa+\sigma}}.
\]
Luego, siguiendo el mismo razonamiento para llegar a la ecuaci\'on (\ref{eq:amplitud}), tenemos que
\[
\dfrac{\left\langle \Delta A\right\rangle}{A_{0}}=\dfrac{K_{B}T}{8\pi\kappa}\ln{\left\lbrace \dfrac{q_{max}^{2}\kappa + 
\sigma}{q_{min}^{2}\kappa + \sigma}\right\rbrace }
\]
	o equivalentemente,
\begin{equation}
 \dfrac{\left\langle \Delta A\right\rangle}{A_{0}}=\dfrac{K_{B}T}{8\pi\kappa}\ln{\left\lbrace \dfrac{(2\pi/a)^{2} + 
\sigma/\kappa}{(2\pi/L)^{2} + \sigma/\kappa}\right\rbrace}.
\label{eq:dareas}
\end{equation}

La ecuaci\'on (\ref{eq:dareas}) fue deducida para una membrana plana. Sin embargo, es igualmente v\'alida en el caso de la 
ves\'icula aspirada mediante micropipetas \cite{helfrich2, safran1}. Ahora bien, en dichos experimentos es posible observar como 
se modifica el \'area aparente de la ves\'icula inicialmente libre de tensiones cuando se le aplica una tensi\'on peque\~na. En 
todo el proceso, tanto antes como despu\'es de aplicar la tensi\'on (siempre que esta \'ultima sea peque\~na), el \'area real de 
la membrana, $A_{real}$, se mantiene constante, como se explic\'o en la Figura \ref{fig:micropipetas}. En lo que sigue denotamos 
por $A_{\sigma}$ el \'area aparente de la membrana correspondiente a la tensi\'on $\sigma$ y definimos la siguiente cantidad:
\begin{equation}
	 \Delta A_{\sigma}=A_{real}-A_{\sigma}.
\label{eq:contension}
\end{equation}
	
Apliquemos ahora la ecuaci\'on (\ref{eq:dareas}) para dos casos dife\-rentes:
\begin{enumerate}
 \item Si la porci\'on de la ves\'icula no est\'a sujeta a tensi\'on externa ($\sigma=0$ en la ecuaci\'on \ref{eq:dareas}) tenemos que: 
	\begin{equation}
	 	\dfrac{\Delta A_{0}}{A_{0}}=\dfrac{K_{B}T}{8\pi\kappa}\ln{\left\lbrace \dfrac{A_{0}}{a^{2}}\right\rbrace}.
	\label{eq:asint}
	\end{equation}
\item Si la porci\'on de la ves\'icula de inter\'es est\'a sujeta a una tensi\'on $\sigma$ tal que 
	$(2\pi/L)^{2}<\sigma/\kappa<(2\pi/a)^{2}$, entonces tenemos que:
	\begin{equation}
	\dfrac{\Delta A_{\sigma}}{A_{\sigma}}=\dfrac{K_{B}T}{8\pi\kappa}\ln{\left\lbrace 
	 \dfrac{4 \pi^{2}\kappa}{a^{2}\sigma}\right\rbrace }.
	\label{eq:acont}
	\end{equation}
\end{enumerate}

Ahora, n\'otese primero que 
\[
 A_{real}=\Delta A_{0}+A_{0}=\Delta A_{\sigma}+A_{\sigma},
\]
y, por lo tanto,
\[
 \dfrac{A_{\sigma}}{A_{0}}=\dfrac{\Delta A_{0}}{A_{0}}-\dfrac{\Delta A_{\sigma}}{A_{0}}+1.
\]
As\'i, sustituyendo en esta \'ultima relaci\'on las ecuaciones (\ref{eq:asint}) y (\ref{eq:acont}), se sigue que
\[
 	\dfrac{A_{\sigma}}{A_{0}}=1+\beta\ln\left\lbrace\dfrac{A_{0}}{a^{2}} \right\rbrace -
\beta\dfrac{A_{\sigma}}{A_{0}}\ln\left\lbrace\dfrac{4 \pi^{2}\kappa}{a^{2}\sigma} \right\rbrace,
\]
donde $\beta\equiv\frac{K_{B}T}{8\pi\kappa}$.
   Luego, factorizando el t\'ermino $\frac{A_{\sigma}}{A_{0}}$ y simplificando, obtenemos finalmente que
\begin{equation}
 	\dfrac{A_{\sigma}}{A_{0}}=\dfrac{1+\beta\ln\left\lbrace\dfrac{A_{0}}{a^{2}} \right\rbrace}
{1+\beta\ln\left\lbrace\dfrac{4 \pi^{2}\kappa}{a^{2}\sigma} \right\rbrace}.
\label{eq:rareas}
\end{equation}
En el experimento es posible medir directamente las \'areas aparentes $A_{0}$ y $A_{\sigma}$, y calcular la deformaci\'on 
$\alpha \equiv \frac{A_{\sigma}-A_{0}}{A_{0}}=\frac{A_{\sigma}}{A_{0}}-1$ que, en nuestro caso, en virtud de (\ref{eq:rareas}) 
puede calcularse mediante
\begin{equation}
	 \alpha=\dfrac{1+\beta\ln\left\lbrace\dfrac{A_{0}}{a^{2}} \right\rbrace}
{1+\beta\ln\left\lbrace\dfrac{4 \pi^{2}\kappa}{a^{2}\sigma} \right\rbrace} - 1.	
\label{eq:deformacionf}
\end{equation}

En particular, para el caso en que $\beta\ll 1$, condici\'on que se cumple al consi\-derar membranas r\'igidas como las formadas 
por fosfol\'ipidos, la deformaci\'on puede aproximarse como
\[
 \alpha\approx\left[1+\beta\ln\left\lbrace\dfrac{A_{0}}{a^{2}} \right\rbrace\right] \left[ 1-\beta\ln\left\lbrace\dfrac{4 \pi^{2}
\kappa}{a^{2}\sigma} \right\rbrace\right]  - 1,
\]
es decir,
\begin{equation}
 \alpha\approx\dfrac{K_{B} T}{4\pi\kappa}\ln\left\lbrace\dfrac{A_{0}\sigma}{4 \pi^{2}\kappa} \right\rbrace.
\label{eq:deformacion}
\end{equation}

\begin{figure}
 \centering
 \includegraphics[width=0.40\textwidth]{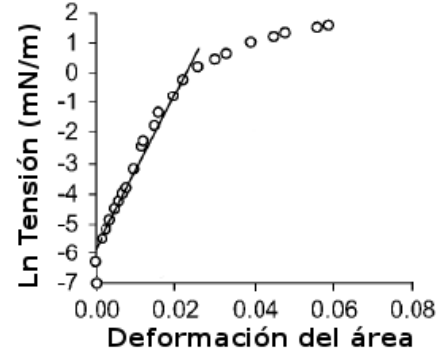}
 \caption{{\small Gr\'afica de tensi\'on-deformaci\'on para una ves\'icula de SOPC en soluci\'on de metanol/agua. 
Los puntos obtenidos de trazar el logaritmo natural de la tensi\'on, $\sigma$, contra la deformaci\'on de \'area, $\alpha$, 
tienen un comportamiento lineal en el reg\'imen de baja tensi\'on 
(0.001-0.5 mN/m). Gr\'afica tomada de \cite{hung}.}}
 \label{fig:graficapip2}
\end{figure}

Esta \'ultima ecuaci\'on permite calcular el cambio de \'area aparente cuando se succiona una ves\'icula con presiones bajas mediante 
la t\'ecnica de aspiraci\'on con micropipetas. Como hemos dicho, en este experimento se inhiben mec\'anicamente las fluctuaciones, y 
el \'area de la ves\'icula correspondiente a la proyecci\'on dentro de la micropipeta es igual al exceso inicial de \'area respecto 
a la forma esf\'erica de la ves\'icula libre. Si se grafica la tensi\'on aplicada en funci\'on del cambio de \'area (ecuaci\'on 
\ref{eq:deformacion}), de la pendiente de la curva es posible extraer $\kappa$. En la Figura \ref{fig:graficapip2} se muestra una 
gr\'afica experimental (tomada de la referencia \cite{hung}) del logaritmo de $\sigma$ vs. $\alpha$, encontrandose que efectivamente 
hay un reg\'imen lineal para bajas tensiones. El ajuste de la pendiente de dicha recta en los experimentos mencionados da una 
medici\'on experimental directa del m\'odulo de curvatura media, $\kappa$. El intervalo t\'ipico para el fosfol\'ipido SOPC es de 
$\kappa\approx 34-41K_{B}T$ \cite{boal, antonietti}. 

\section{\label{sec:fases}Fases formadas por bicapas}

En la secci\'on \ref{sec:bicapas} mencionamos que las mol\'eculas anfif\'ilicas en soluci\'on acuosa forman diferentes estructuras 
debido a su naturaleza dual. Las estructuras b\'asicas son las micelas y las membranas. Estas \'ultimas pueden existir en diferentes 
configuraciones, entre las cuales la m\'as com\'un es la fase lamelar. Sin embargo, las membranas tambi\'en se agrupan en arreglos 
desordenados como la fase esponja. 

\subsection{\label{sec:transicion}Transici\'on entre la fase lamelar y la fase esponja}

\begin{figure}
	\centering
		\includegraphics[width=0.45\textwidth]{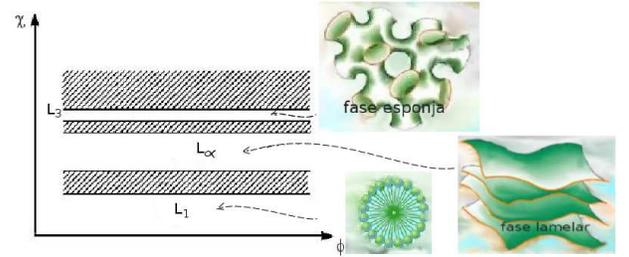}
	\caption{{\small Diagrama de fase esquem\'atico en t\'erminos de los par\'ametros $\chi$ y $\phi$ que resume las 
caracter\'isticas principales de las fases en sistemas de surfactante y agua. Gr\'afica tomada de \cite{porte}.}}
	\label{fig:gfasesll}
\end{figure}

Es un hecho experimental, verificado en muchos sistemas de surfactantes, que las fases lamelar y esponja aparecen en 
regiones vecinas del diagrama de fases. De hecho, hay un patr\'on que se repite en muchos sistemas: la aparici\'on sucesiva de 
las fases micelar, lamelar y esponja al variar un par\'ametro fisicoqu\'imico $\chi$, el cual puede ser la temperatura, la 
concentraci\'on de tensoactivo, la salinidad, la concentraci\'on de un aditivo, etc. Usualmente, las tres fases observadas 
dependen de $\chi$ como se muestra en la Figura \ref{fig:gfasesll}. Para valores peque\~nos de $\chi$ se observa la fase micelar. 
Posteriormente aparecen las fases lamelar y esponja conforme aumenta $\chi$.

Esta sucesi\'on de fases intrig\'o durante mucho tiempo a la comunidad que trabaja en sistemas coloidales autoasociativos. En 
particular, los investigadores se pregun\-taban c\'omo es posible que un cambio tan sutil en las condiciones fisicoqu\'imicas 
(variaci\'on del par\'ametro $\chi$) lleve a una reorganizaci\'on tan dr\'astica de la estructura membranal. Uno de los grandes 
logros de la teor\'ia de Helfrich consiste en que este fen\'omeno puede ser explicado en t\'erminos de las propiedades el\'asticas 
de membranas. A continuaci\'on veremos un modelo sencillo que explica la transici\'on lamelar-esponja en estos sistemas.

	Considerando la energ\'ia de una bicapa de mol\'eculas anfif\'ilicas es posible construir un modelo sencillo que explique por 
qu\'e una fase lamelar se transforma en una fase esponja al variar el par\'ametro $\chi$. Para esto se modela a la fase esponja como 
una superficie peri\'odica m\'inima de Schwarz \cite{porte}, es decir, una superficie con curvatura media $H$ id\'enticamente cero. 
A\'un cuando los experimentos de microscop\'ia electr\'onica indican que la fase esponja es desordenada \cite{porte} (a diferencia de 
la fase c\'ubica que es peri\'odica), esta simplificaci\'on ayuda a visualizar dos caracter\'isticas esenciales de la fase: 1) En 
cualquier parte de la estructura los dos radios principales de curvatura tienen similar magnitud pero signos opuestos. 2) La curvatura
 gaussiana tiene en cualquier parte una magnitud grande comparada con la curvatura media y tiene signo negativo.

La primer caracter\'istica permite despreciar a la curvatura media (de hecho, esta es id\'enticamente cero cuando la fase es una 
superficie peri\'odica m\'inima). Bajo dichas condiciones, las contribuciones de la curvatura media a la energ\'ia libre de Helfrich 
para la fase lamelar y para la esponja son b\'asicamente iguales. Por lo tanto, la principal diferencia entre estas fases es debida 
al t\'ermino gaussiano, como se observa en la ecuaci\'on (\ref{eq:helfricht}).

Por otra parte, es posible deducir a partir del teorema de Gauss-Bonnet que la diferencia en la energ\'ia de Helfrich por unidad de 
\'area, ecuaci\'on (\ref{eq:helfricht}), entre las fases lamelar y esponja es del orden de 
$E_{lamelar} - E_{esponja} = -4p\pi\bar{\kappa}$, como se sigue de la ecuaci\'on (\ref{eq:gbonet}). En base a esto podemos ver qu\'e 
valores del m\'odulo el\'astico gaussiano estabilizan la fase lamelar y cu\'ales la fase esponja. El criterio a seguir es la 
minimizaci\'on de la energ\'ia el\'astica. As\'i, se espera que el valor de $\bar{\kappa}$ que estabiliza la fase lamelar sea negativo, 
pues en ese caso, esta fase tendr\'ia menos energ\'ia el\'astica que la esponja. En cambio, para estabilizar la fase esponja, 
$\bar{\kappa}$ deber\'a ser positiva.

Para entender la transici\'on lamelar-esponja, vamos a describir un modelo que predice c\'omo el par\'ametro fisicoqu\'imico $\chi$ 
afecta el valor del m\'odulo el\'astico gaussiano $\bar{\kappa}$. Para ello vamos a analizar la energ\'ia libre de Helfrich de una 
bicapa partiendo de la energ\'ia de cada una de las monocapas que la constituyen.

Consideramos que las monocapas no son sim\'etricas respecto a la mitad de su superficie, es decir, presentan curvatura espont\'anea 
$c_{0}$. As\'i, la energ\'ia de curvatura para cada monocapa puede escribirse como:
\begin{equation}
 	f_{m}=\dfrac{1}{2}\kappa_{m}\left(H-c_{0}\right)^{2}+\tilde{\kappa}_{m}K. 
\label{eq:energiam}
\end{equation}
	Puesto que en el proceso de pasar de una bicapa plana a una curvada no se involucra cambio de topolog\'ia, es posible 
abandonar el t\'ermino gaussiano al hacer los c\'alculos y reincorporarlo en la expresi\'on final. Para evitar confusiones en este 
c\'alculo, usamos los sub\'indices $m$ y $b$ se\~nalando que estamos hablando de una monocapa o una bicapa, respectivamente. Por 
ejemplo, $\kappa_{m}$ es el m\'odulo de curvatura media de una monocapa y $\kappa_{b}$ el m\'odulo de curvatura media de una bicapa.

\begin{figure}
	\centering
	\includegraphics[width=0.35\textwidth]{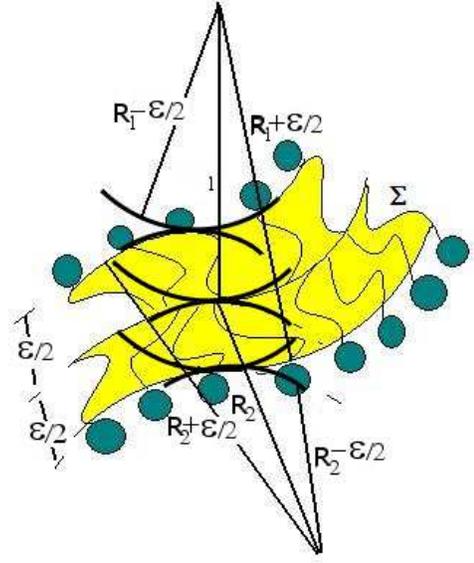}
\caption{{\small Estructura esquem\'atica de una bicapa. La superficie incompresible de cada monocapa permanece a una 
distancia $\epsilon/2$ de la mitad de la superficie $\Sigma$.}}
	\label{fig:monocapa}
\end{figure}

	Cuando las monocapas est\'an unidas frente a frente formando la bicapa, cada una est\'a en un estado que no corresponde a su 
curvatura de equilibrio. La presencia de la otra monocapa hace que necesariamente alguna de las dos no se pueda plegar en la 
direcci\'on requerida por la curvatura espont\'anea. Para una monocapa la curvatura espont\'anea es $1/R_{1}=c_{0}$, mientras que 
para la otra es $-1/R_{2}= - c_{0}$ (ver Figura \ref{fig:monocapa}). Los radios de curvatura que tiene cada monocapa definen su 
energ\'ia el\'astica seg\'un la ecuaci\'on (\ref{eq:energiam}).  La energ\'ia libre de la bicapa puede ser formulada sumando las 
contribuciones de las monocapas.  As\'i, sustituyendo la curvatura media en funci\'on de los radios principales de curvatura de cada 
monocapa a una distancia  $\epsilon/2$, medida desde la mitad de la superficie (Figura \ref{fig:monocapa}) en  la ecuaci\'on 
(\ref{eq:energiam}), tenemos que
\begin{eqnarray*}
&f_{b}=&\dfrac{1}{2}\kappa_{m}\left[ \dfrac{1}{2} \left\lbrace \dfrac{1}{R_{1}+\frac{\epsilon}{2}}+\dfrac{1}{R_{2}+
\frac{\epsilon}{2}} \right\rbrace -c_{0}\right]^{2}\\
& & +\dfrac{1}{2}\kappa_{m}\left[ \dfrac{1}{2} \left\lbrace \dfrac{1}{R_{1}-\frac{\epsilon}{2}}+\dfrac{1}{R_{2}-
\frac{\epsilon}{2}} \right\rbrace +c_{0}\right]^{2}. 
\end{eqnarray*}
	Haciendo un desarrollo en serie de Taylor a segundo orden alrededor de $\epsilon/2$ tenemos
\begin{eqnarray*}
&f_{b}\approx&\dfrac{\kappa_{m}}{8}\left[ \dfrac{1}{R_{1}}-\dfrac{\epsilon}{2R^{2}_{1}}+\dfrac{1}{R_{2}}-
\frac{\epsilon}{2R^{2}_{2}} -2c_{0}\right]^{2}\\
& & +\dfrac{\kappa_{m}}{8}\left[\dfrac{1}{R_{1}}+\dfrac{\epsilon}{2R^{2}_{1}}+\dfrac{1}{R_{2}}+
\frac{\epsilon}{2R^{2}_{2}} +2c_{0}\right]^{2}. 
\end{eqnarray*}
	Desarrollando los par\'entesis y despreciando los t\'erminos de tercer orden y superior en los radios principales de curvatura
\begin{eqnarray*}
&f_{b}\approx&\dfrac{1}{2}\kappa_{m}\left[ \dfrac{1}{2 R_{1}^{2}}+\dfrac{1}{2 R_{2}^{2}} +\dfrac{1}{R_{1}R_{2}}+
\left(\frac{1}{R^{2}_{1}}+\dfrac{1}{R^{2}_{2}}\right)c_{0}\epsilon\right]\\
& & +\kappa_{m}c_{0}^{2} 
\end{eqnarray*}
	lo que puede reescribirse como
\begin{eqnarray*}
&f_{b}=&\dfrac{1}{4}\kappa_{m}\left\lbrace  \dfrac{1}{R_{1}}+\dfrac{1}{R_{2}}\right\rbrace ^{2}\\
& &+\dfrac{1}{2}\kappa_{m}\left(c_{0}\epsilon
\left[ \left(\frac{1}{R_{1}}+\dfrac{1}{R_{2}}\right)^{2} -\dfrac{2}{R_{1}R_{2}}\right]+2c_{0}^{2}\right). 
\end{eqnarray*}
Considerando ahora las definiciones de las curvaturas media y gaussiana en funci\'on de los radios principales de curvatura, tenemos que
\[
 	f_{b}=\dfrac{1}{2}(2\kappa_{m})H^{2}+ 2 H^{2}\epsilon c_{0}\kappa_{m}-K\epsilon c_{0}\kappa_{m}+c_{0}^{2}\kappa_{m}.
\]
	Puesto que $c_{0}^{2}\kappa_{m}$ es una constante independiente de la forma que adquiere la membrana y 
$2 \epsilon H^{2}c_{0}\kappa_{m}$ es despreciable en comparaci\'on con $H^{2}\kappa_{m}$, es posible despreciar estos t\'erminos, 
con lo que se obtiene
\begin{equation}
 	f_{b}=\dfrac{1}{2}(2\kappa_{m})H^{2}+\left(-\epsilon c_{0}\kappa_{m}\right)K.
\label{eq:energiabi}
\end{equation}
Comparando el t\'ermino que involucra la curvatura gaussiana en esta expresi\'on con el de la ecuaci\'on (\ref{eq:helfrich}), se puede 
concluir que 
\begin{equation}
 	\bar{\kappa}_{b}=-\epsilon c_{0}\kappa_{m}.
\label{eq:kappabi}
\end{equation}

Esta ecuaci\'on es importante pues relaciona una constante el\'astica de la membrana ($\bar{\kappa}_{b}$) con propiedades de las 
monocapas que la constituyen: su curvatura espont\'anea ($c_{0}$) y su m\'odulo el\'astico de curvatura media ($\kappa_{m}$). 

Esto permite entender la transici\'on lamelar-esponja de la siguiente manera. Al variar las condiciones fisicoqu\'imicas (par\'ametro 
$\chi$) en la direcci\'on apropiada, las propiedades de las monocapas cambian en consecuencia, por ejemplo, su curvatura espont\'anea. 
Esto ocasiona que el m\'odulo de curvatura gaussiano de la membrana cambie (ecuaci\'on \ref{eq:kappabi}), lo que, si el cambio tiene 
los valores apropiados, se refleja en un cambio de topolog\'ia de membranas abiertas (fase lamelar) a membranas en forma de silla 
de montar a caballo (fase esponja).

Usando la ecuaci\'on (\ref{eq:kappabi}) podemos ver que, en la aproximaci\'on en que la fase esponja es modelada como una superficie 
m\'inima de curvatura media nula, la energ\'ia libre de curvatura tiene la forma 
\[
 	f=-\epsilon c_{0}\kappa_{m}K
\]
y por el teorema de Gauss-Bonnet, ecuaci\'on (\ref{eq:gbonet}), la energ\'ia sobre toda la membrana est\'a dada por
\[
 	F=-4\pi\epsilon c_{0}\kappa_{m}(1-p).
\]

En esta ecuaci\'on vemos que el g\'enero de la superficie, $p$, es tambi\'en un factor que influye en la energ\'ia el\'astica de 
la membrana.

En esta expresi\'on la \'unica constante que puede cambiar de signo es la curvatura espont\'anea de la monocapa. De donde es 
claro que si $c_{0}<0$ la energ\'ia m\'inima corresponde a una superficie cuyo g\'enero es mayor que uno (pues $p$ mayor que uno 
resulta en energ\'ias negativas). Es decir, la energ\'ia el\'astica favorece la fase esponja. Sin embargo, cuando $c_{0}>0$ la 
energ\'ia m\'inima corresponde a una superficie sin agujeros ($p=0$) y como consecuencia la fase lamelar es 
mayormente favorecida. Vemos pues que este modelo sencillo permite entender, al menos cualitativamene, cu\'ando la fase lamelar o 
la fase esponja son estables en t\'erminos de la energ\'ia el\'astica de las membranas.

\subsection{\label{sec:esterica}Interacci\'on est\'erica de Helfrich}

En esta secci\'on vamos a comentar brevemente una consecuencia m\'as de la elasticidad de las membranas. Como veremos, este efecto 
tiene que ver con las interacciones que estabilizan arreglos de membranas como la fase lamelar o la fase esponja.

Para ejemplificar, consideremos una fase lamelar, la cual presenta orden a una escala mucho mayor que el espesor de las membranas. 
La distancia media entre bicapas en esta fase est\'a determinada por las interacciones entre ellas. Por ejemplo, las fuerzas de van 
der Waals, que son siempre atractivas y de corto alcance, tienden a hacer que las membranas se adhieran entre s\'i y que la fase 
lamelar se desestabilice y colapse. En la pr\'actica esto no sucede debido al efecto estabilizador de fuerzas repulsivas, como en el 
caso de las membranas cargadas, donde la interacci\'on repulsiva es electrost\'atica.

Sin embargo, veremos que en sistemas sin carga tam\-bi\'en es posible estabilizar una fase lamelar con distancias entre membranas muy 
grandes, del orden de 1000 \AA. En un principio no se entend\'ia cual era la fuerza repulsiva que lograba equilibrar a las 
interacciones de van der Waals en este caso de ausencia de fuerzas electrost\'aticas. Ahora sabemos que esta fuerza tiene su origen 
en las fluctuaciones de forma de la membrana.

Cualitativamente, este efecto repulsivo puede entenderse de la siguiente manera. Cuando las membranas de un sistema son muy flexibles, 
las fluctuaciones de curvatura tienen una amplitud muy grande (inversamente proporcional al m\'odulo de curvatura media, que es 
peque\~no). Sin embargo, el confinamiento de cada membrana por sus vecinas en la fase lamelar reduce efectivamente el espacio 
configuracional que cada bicapa puede explorar con sus fluctuaciones, tendiendo a reducir la entrop\'ia del sistema. Es precisamente 
esta reducci\'on de entrop\'ia, efecto no favorecido en t\'erminos de la energ\'ia libre, la que da lugar a la fuerza repulsiva 
conocida como interacci\'on est\'erica de Helfrich. El sistema alcanza un equilibrio, donde la energ\'ia libre es m\'inima.

\begin{figure}
	\centering
		\includegraphics[width=0.15\textwidth]{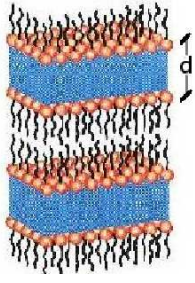}
	\caption{{\small Vista esquem\'atica de una fase lamelar formada por arreglos peri\'odicos de membranas surfactantes.}}
	\label{fig:lamelar}
\end{figure}

El efecto directo del confinamiento es mejor modelado como una constricci\'on a las posibles conformaciones de la membrana. Si la 
distancia promedio entre las membranas es $d$, (Figura \ref{fig:lamelar}), de acuerdo con la ecuaci\'on (\ref{eq:aplituda}), el 
confinamiento ser\'a sentido por una membrana cuando su tama\~no sea del orden de $L_{d}$, tal que
\[
 	\dfrac{K_{B}T}{\kappa}L^{2}_{d}\approx d^{2}.
\]
Podemos ahora pensar en la bicapa dividida en bloques de tama\~no lateral $L_{d}\times L_{d}$ y suponer que cada uno de los bloques 
no es afectado por las interacciones entre las membranas vecinas. As\'i, el confinamiento no restringe las posibles conformaciones 
internas dentro de un bloque. Sin embargo, s\'i restringe el n\'umero de posibles confi\-guraciones de los centros de masas de todos 
los bloques en la bicapa; esto es, restringe los posibles valores de $N_{d} = \left( L / L_{d} \right)^{2}$ coordenadas 
configuracionales. Hagamos ahora la analog\'ia con un gas ideal tridimensional de $N$ part\'iculas, cuya entrop\'ia es 
proporcional a $3 N K_{B}$, donde $K_{B}$ es la constante de Boltzmann. Si un n\'umero $N_{d}$ de las coordenadas configuracionales 
fueran restringidas, la p\'erdida de entrop\'ia ser\'a proporcional a $- N_{d} K_{B} $. Aplicando el mismo criterio a nuestro 
sistema encontramos que la energ\'ia libre debida a la p\'erdida de entrop\'ia por el confinamiento es
\[
 F = - T S \approx K_{B} T N_{d} = K_{B} T \left( \dfrac{L}{L_{d}} \right)^{2} \approx \dfrac{1}{\kappa} 
\left(\dfrac{L}{d}K_{B} T \right)^{2}.
\]
Por consiguiente, la energ\'ia libre de las interacciones est\'ericas por unidad de \'area de una membrana en la fase lamelar 
puede escribirse como
\[
 	\dfrac{F}{A} \approx \dfrac{1}{\kappa} \left( \dfrac{K_{B} T}{d} \right)^{2}.
\]  
Otras derivaciones de la f\'ormula obtenida pueden encontrarse en la literatura \cite{helfrich-esterica,deserno}. Esta energ\'ia 
corresponde a una interacci\'on repulsiva entre bicapas, que var\'ia como el inverso del cuadrado de la distancia de separaci\'on 
entre ellas. Cuando \'esta es combinada con otro tipo de interacci\'on como la de van der Waals o la electrost\'atica, estas 
interacciones determinan la periodicidad de la fase lamelar. Este resultado fue derivado en 1978 por Helfrich \cite{helfrich-esterica}, 
mostrando que las propiedades de la fase lamelar son determinadas en buena medida por las propiedades el\'asticas de las membranas.

\section{Conclusiones}
En este art\'iculo presentamos una derivaci\'on de la energ\'ia de Helfrich, la cual describe las propiedades el\'asticas de membranas. 
Seg\'un este modelo, toda membrana tiene dos m\'odulos el\'asticos: el de curvatura media y el de curvatura gaussiana. El primero de 
ellos est\'a relacionado con la rigidez de la membrana y determina que tanto se deforma, sin cambiar de topolog\'ia, debido a las 
fluctuaciones t\'ermicas. El m\'odulo el\'astico de curvatura gaussiana est\'a relacionado al tipo de topolog\'ia que adquiere una 
membrana. En el texto describimos un m\'etodo para medir el m\'odulo de curvatura media de una membrana, basado en la manipulaci\'on 
de ves\'iculas con micropipetas.
Hemos visto tambi\'en que la elasticidad de las membranas da lugar a consecuencias f\'isicas importantes, como la transici\'on entre 
una fase lamelar y otra esponja, o la existencia de una interacci\'on est\'erica entre membranas.
Las propiedades el\'asticas de membranas juegan un papel importante en la forma que adquieren  las membranas biol\'ogicas en la 
c\'elula. Diversos grupos de investigaci\'on han explotado estos conceptos para predecir con cierto \'exito la forma tanto de 
c\'elulas como los gl\'obulos rojos \cite{PhysRA1990}, o de organelos como el complejo de Golgi \cite{golgi}. Dejamos para un 
trabajo posterior una descripci\'on de algunos de estos resultados.
Cabe decir que el modelo presentado es relativamente simple. De hecho, s\'olo considera membranas fluidas con peque\~nas 
fluctuaciones, despreciando efectos como el citoesqueleto de la c\'elula, las deformaciones laterales, las prote\'inas  
transmembranales y dem\'as complejidades presentadas en las membranas biol\'ogicas. Sin embargo, en la literatura hay extensiones 
del modelo, las cuales rescatan como l\'imite el modelo de Helfrich \cite{cTu0, cTu, cTu1}. 
Sin embargo, pese a sus limitantes, el modelo de Helfrich ha sido muy exitoso y es una referencia fundamental en el estudio de 
membranas biol\'ogicas.

\begin{acknowledgments}
Agradecemos a Gerardo Paredes y Carlos Luna, estudiantes de la Universidad de Sonora, por compartir con nosotros sus gr\'aficas 
y resultados experimentales de manipulaci\'on con micropipetas. LMB agradece a Conacyt por la beca de maestr\'ia otorgada.
\end{acknowledgments}

\appendix

\section{\label{sec:equiparticion}El teorema de equipartici\'on de la energ\'ia} 

Consideremos un sistema compuesto por un gran n\'umero de part\'iculas $N$  (que, por simplicidad, supondremos tienen id\'entica 
masa $m$). El estado microsc\'opico del sistema est\'a caracterizado por las $6N$ coordenadas de las posiciones 
($\vec{q_{1}},\vec{q_{2}},... \vec{q_{N}}$)  y los momentos ($\vec{p_{1}},\vec{p_{2}},... \vec{p_{N}}$) de las part\'iculas. 
Supongamos que el sistema est\'a en contacto con un ba\~no t\'ermico a temperatura $T$ con el cu\'al intercambia energ\'ia (en 
forma de calor) pero no volumen ni part\'iculas. Uno de los teoremas fundamentales de la mec\'anica estad\'istica cl\'asica establece 
que la densidad de probabilidad de que el sistema se encuentre en un microestado particular con energ\'ia 
$E = H \left( \vec{q_{1}},... \vec{q_{N}};\vec{p_{1}},... \vec{p_{N}} \right)$ est\'a dada por:
\begin{equation}
 P \left[ \vec{q_{1}},... \vec{q_{N}};\vec{p_{1}},... \vec{p_{N}} \right] = \dfrac{e^{-\beta H 
 \left( \vec{q_{1}},... \vec{q_{N}};\vec{p_{1}},... \vec{p_{N}} \right)}}{\int{d^{N}\vec{q}} \int{d^{N}\vec{p} 
 e^{-\beta H \left( \vec{q_{1}},... \vec{q_{N}};\vec{p_{1}},... \vec{p_{N}} \right)}} },
\label{eq:canonica}
\end{equation}
donde $\beta= \left( K_{B} T \right)^{-1}$ y $K_{B}$ es la constante de Boltzmann. Esta es la llamada ``distribuci\'on can\'onica''. 
La energ\'ia cin\'etica de la i-\'esima part\'icula es $p^{2}_{i} / \left( 2 m\right)$, por lo que es f\'acil ver de la ecuaci\'on 
(\ref{eq:canonica}) que la energ\'ia cin\'etica promedio de cada part\'icula es:
\[
 	\left< \dfrac{p^{2}_{i}}{2 m} \right> = \dfrac{\int{d^{3} \vec{p} \left[ p^{2}_{i} / \left( 2 m\right) \right] 
e^{- \beta \dfrac{p^{2}_{i}}{2 m}} }}{\int{d^{3} \vec{p} e^{- \beta \dfrac{p^{2}_{i}}{2 m}}}} = \dfrac{3}{2} K_{B} T
\] 
De aqu\'i se sigue que la energ\'ia promedio correspondiente a cada coordenada de momento es $\left( K_{B} T \right) /2$. 
Supongamos ahora que nuestro sistema particular consiste en un conjunto de osciladores arm\'onicos desacoplados, de manera que 
la energ\'ia correspondiente a cada coordenada (o ``grado de libertad'') tiene la forma: $k x^{2}_{i} /2$, con $k$ una constante. 
Utilizando nuevamente la ecuaci\'on (\ref{eq:canonica}) es f\'acil ver que:
\[
 \left< \dfrac{k x^{2}_{i}}{2}  \right> = \dfrac{\int{d x_{i} \left[ k x^{2}_{i} / 2 \right] 
e^{- \beta \dfrac{k x^{2}_{i}}{2}} }}{\int{d x_{i} e^{- \beta \dfrac{k x^{2}_{i}}{2}}}} = \dfrac{1}{2} K_{B} T.
\]
Concluimos que cuando la energ\'ia es cuadr\'atica en alguna de las coordenadas (ya sea de posici\'on o de momento) obtendremos que 
la energ\'ia promedio correspondiente a ese grado de libertad es igual a $\left( K_{B} T \right) /2$. Este resultado se conoce como 
el Teorema de Equipartici\'on de la Energ\'ia.

Otro resultado que se sigue de la distribuci\'on can\'onica y de que las coordenadas $x_{i}$ est\'an desacopladas en la expresi\'on 
para la energ\'ia, es que dos coordenadas diferentes, $x_{i}$ y $x_{j}$, son estadisticamente independientes. En efecto,
\begin{equation}
 \left< x_{i} x_{j} \right> = \dfrac{\int{d x_{i}  x_{i}  e^{- \beta \dfrac{k x^{2}_{i}}{2}} }\int{d x_{j}  x_{j}  
e^{- \beta \dfrac{k x^{2}_{j}}{2}} }}{\int{d x_{i} e^{- \beta \dfrac{k x^{2}_{i}}{2}}}\int{d x_{j} e^{- \beta 
\dfrac{k x^{2}_{j}}{2}}}} = 0.
\label{eq:desacoplamiento}
\end{equation}
Regresemos al caso m\'as complejo de la membrana plana con peque\~nas fluctuaciones t\'ermicas de la secci\'on \ref{sec:forma}. De 
acuerdo a la ecuaci\'on (\ref{eq:energiaq}), la energ\'ia de la membrana es cuadr\'atica en las coordenadas (en el espacio de Fourier) 
$h_{\vec{q}}$. A cada uno de los modos de Fourier (desacoplado a los otros modos) corresponde una energ\'ia cuadr\'atica igual a 
$L^{2} \left( \kappa q^{4} + \sigma q^{2} \right) | h_{\vec{q}} |^{2} /2$. En analog\'ia con los casos m\'as simples discutidos arriba, 
la energ\'ia promedio por modo de Fourier ser\'a:
\[
 \left< \dfrac{L^{2} \left( \kappa q^{4} + \sigma q^{2} \right) | h_{\vec{q}} |^{2}}{2} \right> =  \dfrac{1}{2} K_{B} T,
\]
lo que nos lleva directamente a la ecuaci\'on (\ref{eq:equiparticion}).  Del mismo modo, en analog\'ia con el resultado de la 
ecuaci\'on (\ref{eq:desacoplamiento}), obtenemos para el caso de la membrana la ecuaci\'on (\ref{eq:desacopl}).


\begin{thebibliography}{99}
\bibitem{yeagle} P. L. Yeagle, \emph{The Structure of Biological Membranes}, {\bf CRC Press} (2005).

\bibitem{lodish} H. Lodish, A. Berk, S. L. Zipursky, P. Matsudaira, D. Baltimore, and J. Darnell, \emph{Molecular Cell Biology}, 
{\bf Freeman} (2000).

\bibitem{canham} P. B. Canham, \emph {J. Theor. Biol.} {\bf 26}, 61 (1970).


\bibitem{deuling1} H. J. Deuling and W. Helfrich, \emph{J. Phys. (Paris)} {\bf 37}, 1335 (1976).
  

\bibitem{frank} F. C. Frank, \emph{Discuss. Faraday Soc.} {\bf 25}, 19 (1958).


\bibitem{helfrich} W. Helfrich, \emph{ Z. Naturforsch } {\bf 28c}, 693 (1973).
  

\bibitem{singer} S. J. Singer and G. L. Nicolson, \emph{Science} {\bf 175}, 720 (1972).
  

\bibitem{roux} D. Roux, F. Nallet, E. Freyssingeas, G. Porte, P. Bassereau, M. Skouri and J. Marignan, 
\emph {Eur. Phys. Lett.} {\bf 17}, 575 (1992).


\bibitem{caille} A. Caill\'e, C. R. Hebdo, \emph {Acad. Sci. Paris B} {\bf 274}, 981 (1972). 

\bibitem{gennes} P. G. De Gennes, \emph{J. Phys. Colloq. France} {\bf 30}, c4 (1969). 

\bibitem{handbook} Gregor Cevc (ed.), \emph{Phospholipids Handbook}, {\bf Marcel Dekker  Inc.}, (1993).

\bibitem{israelachvili} J. N. Israelachvili, \emph{Intermolecular and Surface Forces}, {\bf Academic Press} (1985).

\bibitem{yang} Zhong-Can Ou-Yang, Ou-Yang Zhong-Can, Ji-Xing Liu, Liu Ji-Xing, Yu-Zhang Xie, Xie Yu-Zhang, \emph{Geometric 
Methods in the Elastic Theory of Membranes in Liquid Crystal Phases}, {\bf World Scientific} (1999).

\bibitem{safran} S. A. Safran \emph{Statistical thermodynamics of surfaces, interfaces, and membranes}, 
{\bf Addison-Wesley} (1994).

\bibitem{kreiszing} E. Kreyszing, \emph{Differential Geometry}, {\bf Dover} (1991).

\bibitem{golgi} J. Derganc, A. Mironov and S. Svetina,  
\emph{Traffic} {\bf 7}, 85 (2006).

\bibitem{fung} Y. C. B. Fung and P. Tomg,  
\emph{Biophys. J.} {\bf 8}, 175 (1968).

\bibitem{evans1} E. Evans and W. Rawicz., \emph{Phys. Rev. Lett.} {\bf 64}, 2094 (1990).


\bibitem{rawicz}  W. Rawicz, K. C. Olbrich, T. McIntosh, D. Needham and E. Evans., \emph{Biophys. J.} {\bf 79}, 328 (2000).


\bibitem{helfrich2} W. Helfrich and R. M. Servuss, \emph{Nuovo Cimento} {\bf 3, 137} (1984). 


\bibitem{ipsen} J. R. Henriksen and J. H. Ipsen, \emph{Eur. Phys. J. E} {\bf 14}, 149 (2004).


\bibitem{hung} V. Ly Hung and L. Longo Marjorie, \emph{Biophys. J.} {\bf 87}, 1013 (2004).


\bibitem{safran1} Scott. T. Milner and S. A. Safran, \emph{Phys. Rev. A} {\bf 36 (9)}, 4371 (1987).
 

\bibitem{boal} D. Boal, \emph{Mechanics of the Cell}, {\bf Cambridge University Press} (2002).

\bibitem{antonietti} M. Antonietti, S. Forster, \emph{Adv. Mater.} {\bf 15}, 1323 (2003).


\bibitem{porte} G. Porte, J. Appell, P. Bassereau, J. Marignan, \emph{J. Phys. France} {\bf 50}, 1335 (1987).


\bibitem{helfrich-esterica} W. Helfrich, \emph{Z. Naturforsch} {\bf 33a}, 305 (1978).


\bibitem{deserno} Deserno M. \emph{Fluid lipid membranes - a primer}\\ http://www.mpip-mainz.mpg.de/$\sim$deserno 
/scripts/IMPRS/IMPRS\_lipidmembranes.pdf

\bibitem{PhysRA1990} Ou-Yang-can, Xie Zhang,\emph{Phys. Rev. A} {\bf 41}, 3381 (1990)

\bibitem{cTu0} Z. C. Tu and Z. C. Ou-Yang, \emph{Phys. Rev. E} {\bf 68}, 061915 (2003).

\bibitem{cTu} Z. C. Tu and Z. C Ou-Yang, \emph{J. Phys. A: Math. Gen.} {\bf 37}, 11407 (2004).

\bibitem{cTu1} Z. C. Tu, L. Q. Ge, J. B. Li, Z. C. Ou-Yang, \emph{Phys. Rev. E} {\bf 72}, 021806 (2005).


\end{thebibliography}
\end{document}